\input harvmac.tex



\def\unlockat{\catcode`\@=11}
\def\lockat{\catcode`\@=12}

\unlockat

\def\newsec#1{\global\advance\secno by1\message{(\the\secno. #1)}
\global\subsecno=0\global\subsubsecno=0\eqnres@t\noindent
{\bf\the\secno. #1}
\writetoca{{\secsym} {#1}}\par\nobreak\medskip\nobreak}
\global\newcount\subsecno \global\subsecno=0
\def\subsec#1{\global\advance\subsecno
by1\message{(\secsym\the\subsecno. #1)}
\ifnum\lastpenalty>9000\else\bigbreak\fi\global\subsubsecno=0
\noindent{\it\secsym\the\subsecno. #1}
\writetoca{\string\quad {\secsym\the\subsecno.} {#1}}
\par\nobreak\medskip\nobreak}
\global\newcount\subsubsecno \global\subsubsecno=0
\def\subsubsec#1{\global\advance\subsubsecno by1
\message{(\secsym\the\subsecno.\the\subsubsecno. #1)}
\ifnum\lastpenalty>9000\else\bigbreak\fi
\noindent\quad{\secsym\the\subsecno.\the\subsubsecno.}{#1}
\writetoca{\string\qquad{\secsym\the\subsecno.\the\subsubsecno.}{#1}}
\par\nobreak\medskip\nobreak}

\def\subsubseclab#1{\DefWarn#1\xdef
#1{\noexpand\hyperref{}{subsubsection}%
{\secsym\the\subsecno.\the\subsubsecno}%
{\secsym\the\subsecno.\the\subsubsecno}}%
\writedef{#1\leftbracket#1}\wrlabeL{#1=#1}}
\lockat

\def\IL{\relax{\rm I\kern-.18em L}}
\def\IH{\relax{\rm I\kern-.18em H}}
\def\IR{\relax{\rm I\kern-.18em R}}
\def\IC{\relax\hbox{$\inbar\kern-.3em{\rm C}$}}
\def\IT{\relax\hbox{$\inbar\kern-.3em{\rm T}$}}
\def\IZ{\relax\ifmmode\mathchoice
{\hbox{\cmss Z\kern-.4em Z}}{\hbox{\cmss Z\kern-.4em Z}}
{\lower.9pt\hbox{\cmsss Z\kern-.4em Z}}
{\lower1.2pt\hbox{\cmsss Z\kern-.4em Z}}\else{\cmss Z\kern-.4em
Z}\fi}


\def\CT {{\cal T}}

\font\manual=manfnt \def\dbend{\lower3.5pt\hbox{\manual\char127}}

\def\IZ{\relax\ifmmode\mathchoice
{\hbox{\cmss Z\kern-.4em Z}}{\hbox{\cmss Z\kern-.4em Z}}
{\lower.9pt\hbox{\cmsss Z\kern-.4em Z}}
{\lower1.2pt\hbox{\cmsss Z\kern-.4em Z}}\else{\cmss Z\kern-.4em
Z}\fi}
\def\half {{1\over 2}}

\def\p{\partial}


\def\IZ{\relax\ifmmode\mathchoice
{\hbox{\cmss Z\kern-.4em Z}}{\hbox{\cmss Z\kern-.4em Z}}
{\lower.9pt\hbox{\cmsss Z\kern-.4em Z}}
{\lower1.2pt\hbox{\cmsss Z\kern-.4em Z}}\else{\cmss Z\kern-.4em
Z}\fi}
\def\IB{\relax{\rm I\kern-.18em B}}
\def\IC{{\relax\hbox{$\inbar\kern-.3em{\rm C}$}}}
\def\ID{\relax{\rm I\kern-.18em D}}
\def\IE{\relax{\rm I\kern-.18em E}}
\def\IF{\relax{\rm I\kern-.18em F}}
\def\IG{\relax\hbox{$\inbar\kern-.3em{\rm G}$}}
\def\IGa{\relax\hbox{${\rm I}\kern-.18em\Gamma$}}
\def\IH{\relax{\rm I\kern-.18em H}}
\def\II{\relax{\rm I\kern-.18em I}}
\def\IK{\relax{\rm I\kern-.18em K}}
\def\IP{\relax{\rm I\kern-.18em P}}
\def\IQ{\relax\hbox{$\inbar\kern-.3em{\rm Q}$}}

\def\inbar{\,\vrule height1.5ex width.4pt depth0pt}

\def\p{\partial}

\font\cmss=cmss10 \font\cmsss=cmss10 at 7pt
\def\IR{\relax{\rm I\kern-.18em R}}


\def\boxit#1{\vbox{\hrule\hbox{\vrule\kern8pt
\vbox{\hbox{\kern8pt}\hbox{\vbox{#1}}\hbox{\kern8pt}}
\kern8pt\vrule}\hrule}}
\def\mathboxit#1{\vbox{\hrule\hbox{\vrule\kern8pt\vbox{\kern8pt
\hbox{$\displaystyle #1$}\kern8pt}\kern8pt\vrule}\hrule}}


\def\inbar{\,\vrule height1.5ex width.4pt depth0pt}

\def\p{\partial}

\font\cmss=cmss10 \font\cmsss=cmss10 at 7pt
\def\IR{\relax{\rm I\kern-.18em R}}



\lref\fs{R. Fintushel and R.J. Stern,
``The blowup formula for Donaldson invariants'',
alg-geom/9405002; Ann. Math. {\bf 143} (1996) 529.}

\lref\gottsche{L. G\"ottsche,
``Modular forms and Donaldson invariants for 4-manifolds with $b_+=1$'',
alg-geom/9506018; J. Am. Math. Soc. {\bf 9} (1996) 827.}

\lref\gottzag{L. G\"ottsche and D. Zagier,
``Jacobi forms and the structure of Donaldson invariants for 4-manifolds with
$b_+=1$'', alg-geom/9612020.}

\lref\mw{G. Moore and E. Witten, 
``Integration over the $u$-plane in Donaldson theory'', 
hep-th/9709193; Adv. Theor. Math. Phys. {\bf 1} (1997) 298.}

\lref\swi{N. Seiberg and E. Witten,
``Electric-magnetic duality, monopole condensation, and confinement in
${\cal N}=2$ supersymmetric Yang-Mills Theory'',
hep-th/9407087; Nucl. Phys. {\bf B 426} (1994) 19.}

\lref\swii{N. Seiberg and E. Witten,
``Monopoles, Duality and Chiral Symmetry Breaking in ${\cal N}=2$
supersymmetric QCD'', 
hep-th/9408099; Nucl. Phys. {\bf B 431} (1994) 484.}

\lref\monopole{E. Witten, 
``Monopoles and four-manifolds'', 
hep-th/9411102; Math. Res. Letters {\bf 1} (1994) 769.}

\lref\witteni{E. Witten, 
``On $S$-duality in Abelian gauge theory'', 
hep-th/9505186; Selecta Mathematica {\bf 1} (1995) 383.}

\lref\wittk{E. Witten, 
``Supersymmetric Yang-Mills theory on a four-manifold'', 
hep-th/9403193; J. Math. Phys. {\bf 35} (1994) 5101.}

\lref\lns{A. Losev, N. Nekrasov, and S. Shatashvili, 
``Issues in topological gauge theory'', 
hep-th/9711108; Nucl. Phys. {\bf B 534}  (1998) 549. 
``Testing Seiberg-Witten solution'', in ``Cargese 1997, Strings, branes and
dualities'', pp. 359--372, hep-th/9801061.}

\lref\mmone{M. Mari\~no and G. Moore, 
``Integrating over the Coulomb branch in ${\cal N}=2$ gauge theory'',
hep-th/9712062.}

\lref\mmtwo{M. Mari\~no and G. Moore, 
``The Donaldson-Witten function for gauge groups of rank larger than one'',
hep-th/9802185; Commun. Math. Phys. {\bf 199} (1998) 25.}

\lref\DoKro{S.K. Donaldson and P.B. Kronheimer,
{\it The Geometry of Four-Manifolds},
Clarendon Press, Oxford, 1990.}

\lref\FrMor{R. Friedman and J.W. Morgan,
{\it Smooth Four-Manifolds and Complex Surfaces},
Springer Verlag, 1991.}

\lref\tqft{E. Witten,
``Topological Quantum Field Theory'',
Commun. Math. Phys. {\bf 117} (1988) 353.}

\lref\naka{T. Nakatsu and K. Takasaki, 
``Whitham-Toda hierarchy and ${\cal N}=2$ supersymmetric Yang-Mills theory'', 
hep-th/9509162; Mod. Phys. Lett. {\bf A11} (1996) 157.}

\lref\ITEP{A. Gorsky, A. Marshakov, A. Mironov and A.Morozov, 
``RG equations from Whitham hierarchy'', 
hep-th/9802007; Nucl. Phys. {\bf B 527} (1998) 690.}

\lref\dpstrong{E. D'Hoker and D.H. Phong, 
``Strong coupling expansions of $SU(N)$ Seiberg-Witten theory'', 
hep-th/9701055; Phys. Lett. {\bf B 397} (1997) 94.}

\lref\takasaki{K. Takasaki,
``Integrable hierarchies and contact terms in $u$-plane integrals of
topologically twisted supersymmetric gauge theories'', 
hep-th/9803217; Int. J. Mod. Phys. {\bf A 14} (1999) 1001.}

\lref\toda{A. Gorsky, I.M. Krichever, A. Marshakov, A. Mironov and A.
Morozov, 
``Integrability and Seiberg-Witten exact solution'', 
hep-th/9505035; Phys. Lett. {\bf B 355} (1995) 466.}

\lref\marwar{E. Martinec and N.P. Warner, 
``Integrable systems and supersymmetric gauge theory'', 
hep-th/9509161; Nucl. Phys. {\bf B 459} (1996) 97.}

\lref\Krichever{I. Krichever, 
``The $\tau$-function of the universal Whitham hierarchy, matrix models and
topological field theories'', 
hep-th/9205110; Comm. Pure Appl. Math. {\bf 47} (1994) 437.}

\lref\itomoro{H. Itoyama and A. Morozov, 
``Integrability and Seiberg--Witten theory: curves and periods'',
hep-th/9511126; Nucl. Phys. {\bf B 477} (1996)  855. 
``Prepotential and the Seiberg--Witten theory'', 
hep-th/9512161; Nucl. Phys. {\bf B 491} (1997) 529.}

\lref\emm{J.D. Edelstein, M. Mari\~no and J. Mas, 
``Whitham hierarchies, instanton corrections, and soft supersymmetry breaking
in ${\cal N}=2$ $SU(N)$ super Yang-Mills theory'', 
hep-th/9805172; Nucl. Phys. {\bf B 541} (1999) 671.}

\lref\marijapon{M. Mari\~no, 
``The uses of Whitham hierarchies'', 
hep-th/9905053; Progr. Theor. Phys. Suppl. {\bf 135} (1999) 29.}

\lref\victoruno{V.M. Buchstaber, V.Z. Enolskii and D.V. Leykin, 
``Kleinian functions, hyperelliptic Jacobians, and applications'', 
in {\it Reviews in Mathematics and Mathematical Physics} {\bf 10:2}, 
eds. S.P. Novikov and I.M. Krichever, London 1997.} 

\lref\baker{H.F. Baker, {\it An introduction to the theory of multiply 
periodic functions}, Cambridge University Press, 1907.} 

\lref\bakertwo{H.F. Baker, {\it Abel's theorem}, Cambridge University Press,
1897.}

\lref\bolzabrios{O. Bolza, 
``Proof of Brioschi's recursion formula for the expansion of the even
$\sigma$-function of two variables'',  
Amer. J. Math. {\bf 21} (1899) 1.} 

\lref\bolzarem{O. Bolza, 
``Remarks concerning the expansions of the hyperelliptic
$\sigma$-functions'',  
Amer. J. Math. {\bf 22} (1900) 101.} 

\lref\bolzadif{ O. Bolza, 
``On the first and second logarithmic derivatives of hyperelliptic
$\sigma$-functions'', Amer. J. Math. {\bf 17} (1895) 11. 
``The partial differential equations for the hyperelliptic $\Theta$ and
$\sigma$-functions'', Amer. J. Math. {\bf 21} (1899) 107.}

\lref\edemas{J.D. Edelstein and J. Mas, 
``Strong coupling expansion and the Seiberg-Witten-Whitham equations'', 
hep-th/9901006; Phys. Lett. {\bf B 452} (1999) 69.}

\lref\reviews{J.D. Edelstein and J. Mas, in {\em Trends in Theoretical
Physics II},  edited by H. Falomir, R.E. Gamboa-Sarav\'{\i} and F.A.
Schaposnik, AIP Conference Proceedings {\bf 484} (1999) 195, hep-th/9902161;
K.~Takasaki, ``Whitham deformations and Tau functions in ${\cal N}=2$
supersymmetric  gauge theories'', hep-th/9905224; R.~Carroll, ``Various
aspects of Whitham times'', hep-th/9905010.}

\lref\egrm{J.D. Edelstein, M. G\'omez--Reino and J. Mas, 
``Instanton corrections in ${\cal N}=2$ supersymmetric theories with
classical gauge groups and fundamental matter hypermultiplets'',
hep-th/9904087; Nucl. Phys. {\bf B 561} (1999) 273.}

\lref\egrmm{J.D. Edelstein, M. G\'omez--Reino, M. Mari\~no and J. Mas,
``${\cal N}=2$ supersymmetric gauge theories with massive hypermultiplets and
the Whitham hierarchy'', 
hep-th/9911115; Nucl. Phys. {\bf B 574} (2000) 587.}

\lref\sunmat{
A.~Hanany and Y.~Oz, Nucl. Phys. 
{\bf B 452} (1995) 283, hep-th/9505075. P.C.~Argyres, M.R.~Plesser and
A.~Shapere, Phys. Rev. Lett. {\bf 75} (1995) 1699, hep-th/9505100; }

\lref\sun{A. Klemm, W. Lerche, S. Theisen and S. Yankielowicz, 
``Simple singularities and ${\cal N}=2$ supersymmetric Yang-Mills theory'', 
hep-th/9411048; Phys. Lett. {\bf B 344} (1995) 169.}

\lref\argfar{P.C. Argyres and A.E. Faraggi, 
``Vacuum  structure and spectrum of ${\cal N}=2$ supersymmetric $SU(n)$ gauge 
theory'', 
hep-th/9411057; Phys. Rev. Lett. {\bf 74} (1995) 3931.}

\lref\Tak{K. Takasaki, 
``Whitham Deformations of Seiberg--Witten Curves for Classical Gauge Groups'',
hep-th/9901120. 
``Whitham Deformations and Tau  Functions in ${\cal N} = 2$ Supersymmetric
Gauge Theories'', 
hep-th/9905224; Prog. Theor. Phys. Suppl. {\bf 135} (1999) 53.}

\lref\fay{J. Fay, 
{\it Theta Functions on Riemann Surfaces}, Lect. Notes on Math. {\bf 352}, 
Springer 1973.}

\lref\mum{D. Mumford, {\it Tata Lectures on Theta I} and {\it II}, Prog. in
Math. {\bf 28} and {\bf 43}, Birkh\"auser, 1983 (84).}

\lref\bakerpaper{H.F. Baker, ``On the hyperelliptic $\sigma$-functions'', 
Amer. J. Math. {\bf 20} (1898) 301. }

\lref\gh{P. Griffiths and J. Harris, {\it Principles of Algebraic Geometry}, 
1976, Wiley.}

\lref\akhi{N. I. Akhiezer, {\it Elements of the theory of 
elliptic functions,} AMS, 1990.}

\lref\zj{P. Di Francesco, P. Ginsparg and J. Zinn-Justin, 
``2D gravity and random matrices'', 
hep-th/9306153; Phys. Rept. {\bf 254} (1995) 1.}

\lref\wittopgrav{E. Witten, ``Two-dimensional topological gravity and 
intersection theory on the moduli space'', Surv. Diff. Geom. {\bf 1} (1991) 
243.}

\lref\belo{E.D. Belokolos, A.I. Bobenko, V.Z. Enolskii, A.R. Its, and 
V.B. Matveev, {\it Algebro-geometric approach to nonlinear 
integrable equations,} Springer-Verlag, 1994.}

\lref\ds{M.R. Douglas and S.H. Shenker, 
``Dynamics of $SU(N)$ supersymmetric gauge theory'', 
hep-th/9503163; Nucl. Phys. {\bf B 447} (1995) 271.}

\lref\fssw{R. Fintushel and R.J. Stern, ``Inmersed spheres in 4-manifolds 
and the inmersed Thom conjecture'', Turkish J. Math. {\bf 19} (1995) 145.}

\lref\munoz{V. Mu\~noz, ``Donaldson invariants of non-simple type 
4-manifolds'', math.DG/9909165.}

\lref\klt{A. Klemm, W. Lerche and S. Theisen, ``Nonperturbative effective 
actions of ${\cal N}=2$ supersymmetric gauge theories'', 
hep-th/9505150; Int. J. Mod. Phys. {\bf A 10} (1996) 1029.}

\lref\egm{J.D. Edelstein, M. G\'omez-Reino and M. Mari\~no, ``Remarks on
twisted theories with matter'', hep-th/0011227; J. High Energy Phys. {\bf
01} (2001) 004.}

\lref\victordos{V.M. Buchstaber, V.Z. Enolskii and D.V. Leykin, 
`` Hyperelliptic Kleinian functions and
applications'', in {\it  Solitons, geometry, and topology: 
on the crossroad}, Amer. Math. Soc. Transl. Ser. 2 {\bf 179} 
(1997) 1.}
\lref\victortres{J.C. Eilbeck, V.Z. Enolskii and D.V. Leykin, 
``On the Kleinian construction of Abelian functions of canonical 
algebraic curves'',  in {\it Symmetries of integrable difference 
equations}, Proceedings of the conference SIDE III, 1998.}

\Title{\vbox{\baselineskip12pt
\hbox{US-FT/11-00 }
\hbox{RUNHETC-2000-22 }
\hbox{hep-th/0006113}
}}
{\vbox{\centerline{Blowup formulae in Donaldson-Witten theory}
\centerline{ }
\centerline{and integrable hierarchies}}
}
\centerline{Jos\'e D. Edelstein$^{a}$\foot{Address after July 15, 2000:
Department of Physics, Harvard University, Cambridge, MA 02138, USA.}, Marta
G\'omez-Reino$^{a}$ and Marcos Mari\~no$^{b}$}

\bigskip
\medskip
{\vbox{\centerline{$^{a}$ \sl Departamento de F\'\i sica de Part\'\i culas}
\centerline{\sl Universidade de Santiago de Compostela}
\vskip2pt
\centerline{\sl E-15706 Santiago de Compostela, Spain}}
\centerline{ \it edels, marta@fpaxp1.usc.es}

\bigskip
\medskip
{\vbox{\centerline{$^{b}$ \sl New High Energy Theory Center}
\centerline{\sl Rutgers University}
\vskip2pt
\centerline{\sl Piscataway, NJ 08855, USA }}
\centerline{ \it marcosm@physics.rutgers.edu }

\bigskip
\bigskip
\noindent

We investigate blowup formulae in Donaldson-Witten theory with gauge group
$SU(N)$, using the theory of hyperelliptic Kleinian functions. We find that
the blowup function is a hyperelliptic $\sigma$-function and we describe an
explicit procedure to expand it in terms of the Casimirs of the  gauge group
up to arbitrary order. As a corollary, we obtain a new expression for the
contact terms  and we show that the correlation functions involving the
exceptional divisor  are governed by the KdV hierarchy. We also show that,
for manifolds  of simple type, the blowup function becomes a $\tau$-function
for a multisoliton solution.

\Date{June, 2000} 

\listtoc \writetoc

\newsec{Introduction}
 
Blowup formulae \fs\ have played an important role in Donaldson-Witten
theory.  First of all, they relate the Donaldson invariants of a manifold
$X$ with those of its blownup $\widehat X$, and they have been a crucial
ingredient in the derivation of explicit expressions for these invariants,
their wall-crossings \gottsche\gottzag, and their structural properties in
the case of non-simple type  manifolds \munoz. Another important aspect of
these formulae is that they give an explicit connection between the
mathematical  and the physical approach to Donaldson invariants. For example,
in the derivation of the blowup formula for $SU(2)$ Donaldson invariants
given in \fs, the elliptic curve of the Seiberg-Witten solution
\swi\swii\  appears in a natural way. Conversely, the result of \fs\ can be 
derived in a very elegant way within the framework of the $u$-plane integral 
of Moore and Witten \mw. 

Donaldson-Witten theory can be generalized to higher rank gauge groups 
using the approach of \mw. A detailed analysis of this theory for $SU(N)$ 
has been made in \mmtwo, and also in \lns\ from a slightly different 
point of view. In particular, one of the results of \mmtwo\lns\ is a blowup
formula for $SU(N)$ Donaldson theory, which is written in terms of theta
functions\foot{In \lns, the blowup formula was also derived in the $SU(2)$ 
case. The general formula for $SU(N)$ is implicit in the results 
presented there, and it was in fact used to obtain expressions for the 
contact terms.}. It was already noticed in \mmtwo\ that the blowup function is
essentially a $\tau$-function of the Toda-KP hierarchy, and reflects the
underlying integrable structure of the low-energy effective theory
\toda\marwar\naka\itomoro. This relation  between blowup functions and
integrable hierarchies has been explored in \takasaki\Tak\marijapon. 

In this paper, we shall analyze in full detail the properties and structure 
of the blowup formulae in $SU(N)$ Donaldson-Witten theory. As we will review 
below, an important aspect of blowup functions is that they must admit an
expansion whose  coefficients are polynomials in the Casimirs of the gauge
group  (equivalently, in the local observables of the corresponding
topological theory). In the case of $SU(2)$, the fact that the expression
for the  blowup formula in terms of theta functions admits such an expansion
is a result of the theory  of elliptic functions, which also provides an
explicit  way of performing the expansion by using elliptic
$\sigma$-functions.
 
In the case of $SU(N)$, it was argued in \mmtwo\ that such an expansion 
should exist on physical grounds, but no recipe was given to perform the
expansion. In this paper  we solve this problem by using the 
hyperelliptic generalization of $\sigma$-functions and the theory of
hyperelliptic Kleinian functions. This theory was developed at the end of
nineteenth  century by Klein, Baker, Bolza, and many others, but has
completely  dropped out of the textbooks. There has been recently some
revival of  this theory in connection with the algebro-geometric approach to 
integrable hierarchies \victoruno\victordos\victortres, and as we 
will show in this paper, the theory of hyperelliptic Kleinian functions is the
right framework to address the properties of the blowup functions in $SU(N)$
Donaldson-Witten theory. For example, the contact terms of two-observables
are deeply related to the blowup function, as it was first realized in \lns.
We will show that the theory of hyperelliptic Kleinian functions gives a
simple expression for these contact terms as periods of certain meromorphic
forms. 

Another interesting aspect of this approach is that it makes possible to 
clarify further the connection to integrable hierarchies. We will show in
detail that the blowup function, after a linear transformation of the coupling
constants appearing in the $u$-plane integral, satisfies the differential
equations of the KdV integrable system. As a corollary, the correlation
functions involving the exceptional divisor on the blownup manifold are
governed by the KdV hierarchy. This gives a formal connection to
two-dimensional topological gravity \wittopgrav. 

As it is well-known, in the $SU(2)$ case the blowup formula has a simple 
structure when the manifold is of simple type, and it corresponds to the  
degeneration of elliptic functions to trigonometric functions \fs. In the 
$SU(N)$ case, the simple type condition corresponds to a maximal degeneration 
of the hyperelliptic curve. These degenerations are well-known in the 
algebro-geometric approach to integrable systems, and correspond to 
multisoliton solutions of the hierarchy (see, for example, \belo\mum). We
will then show that the blowup function of $SU(N)$ becomes a $\tau$-function
for an $(N-1)$-soliton solution of the underlying KdV hierarchy. As a
corollary of this analysis we will give explicit expressions for some 
physical quantities at the ${\cal N}=1$ points of ${\cal N}=2$ $SU(N)$
Yang-Mills theory. 

The paper is organized as follows: in section 2, we review the basic results
on blowup formulae in Donaldson-Witten theory for the gauge group $SU(N)$,
following the results of \mmtwo\lns\marijapon. In section  3, we introduce
Kleinian functions and hyperelliptic $\sigma$-functions  and some of their
properties. In particular, we give a detailed account  of the differential
equations that they satisfy and we present  a systematic way to solve them
for any genus $g$. We apply these results to the Seiberg-Witten curve for
$SU(N)$ in section 4, and we derive some new results on the contact terms of
the twisted theory. We present explicit results for the  expansion of the
blowup functions for $g=2$ and $g=3$. In section 5, we  explain the relation
between the blowup function and the KdV hierarchy. We then consider, in
section 6, the important case of manifolds of simple type, and we compute in
full detail the blowup function at the ${\cal N}=1$  points. Finally, in
section 7 we state our conclusions and prospects for future research in this
subject. 

\newsec{The blowup function in twisted ${\cal N}=2$ super Yang-Mills}

In this section we give a brief review of the blowup formula in twisted 
${\cal N}=2$ Yang-Mills theory. A detailed account can be found in 
\mmtwo\marijapon. 

Twisted ${\cal N}=2$ theories have a finite set of gauge-invariant operators 
called observables which can be understood as BRST cohomology classes. For
$SU(N)$ gauge theories, the simplest observables are the $N-1$  Casimirs of
the gauge group, which give a basis for the ring of local, BRST invariant
 operators of the theory. We will take these observables to be the elementary
symmetric polynomials in the eigenvalues of the  complex scalar field $\phi$
in the ${\cal N}=2$ vector multiplet:
\eqn\casimirs{
{\cal O}_k = S_k(\phi_i)= {1\over k}{\rm Tr \phi}^k + \cdots
\,\,\,\,\,\,\,\,\ k=2, \cdots, N ~.} 
The advantage of these operators is that their vacuum expectation values are
precisely the $u_k$ that parametrize the Coulomb branch of the physical
theory. 

From the above operators one can generate the rest of the observables using
the descent procedure. We will consider only simply connected manifolds, for
simplicity. In this case, the other observables of the theory are associated
to integrals over two-cycles $S$ in the manifold $X$ of differential forms
constructed by acting on the Casimirs with a spin one (descent) operator
$G_\mu$,
\eqn\twoc{
I_k(S) = \int_S G^2 {\cal O}_k ={1 \over k} \int_S {\rm Tr}(\phi^{k-1}F) +
\cdots} 
Here, $F$ is the Yang-Mills field strength. In general, $S$ will be an
arbitrary linear combination of basic two-cycles $S_i$, $i=1, \dots, b_2(X)$,
{\it i.e.} $S=\sum_{i=1}^{b_2(X)} t_i S_i$, therefore
\eqn\sumtwo{
I_k(S)=\sum_{i=1}^{b_2(X)} t_i I_k(S_i) ~.}
In total, we have $(N-1) \cdot b_2(X)$ independent operators $I_k(S_i)$.
The basic problem now is to compute the generating function for correlators 
involving the observables that we have just described, that is:
\eqn\genfun{
Z(p_k,f_k, S)=\Big\langle \exp\bigl[ \sum_k (f_k I_k (S) + p_k {\cal O}_k)
\bigr] \Big\rangle_X ~.}
As it has been explained in \mw\ for $SU(2)$, and generalized in \mmtwo\ to
$SU(N)$, the computation of \genfun\ can be done by using the low-energy
exact solution of ${\cal N}=2$, $SU(N)$ Yang-Mills theory. This solution is
encoded in the hyperelliptic curve describing a genus $g=N-1$ Riemann 
surface $\Sigma_g$ \sun\argfar:
\eqn\hypersw{
y^2 = P_N^2(x) - 4\Lambda^{2N} ~,}
where
\eqn\polychar{ 
P_N (x) = x^{N} - \sum_{k=2}^N u_kx^{N-k}} 
is the characteristic polynomial of $SU(N)$, and $u_k=\langle {\cal O}_k
\rangle$ are the VEVs of the Casimir operators \casimirs. Associated to this
curve there is a meromorphic differential of the second kind (also known as
Seiberg-Witten  differential), with a double pole at infinity, that can be
explicitly written as:
\eqn\swdiff{
dS_{SW} = P_N'(x) {x dx \over y} ~.}
This one-form satisfies the equation:
\eqn\defds{
{\partial dS_{SW} \over \partial u_{k+1} } = dv_k ~,}
where 
\eqn\holo{
dv_k={x^{g-k}dx \over y} ~, \,\,\,\,\,\,\,\,\,\ k=1, \cdots, g ~,}
is a basis of holomorphic differentials for hyperelliptic curves of genus
$g$. Given a symplectic basis of homology cycles $A^i,B_i\in
H_1(\Sigma_g,\IZ)$ one may compute the period integrals of these
differentials:
\eqn\periods{
A^i{_k} ={1\over 2\pi i } \oint_{A^i}dv_k ~, ~~~~~~~~~~~~~ 
B_{ik} ={1\over 2\pi i } \oint_{B_i}dv_k ~.}
(Notice that, in contrast to \ITEP\emm\egrm, we have explicitly included the
$2\pi i $ factors). Using these quantities we can define the period matrix
of $\Sigma_g$ as 
\eqn\gaugeco{
\tau_{ij}=B_{ik} (A^{-1})^k_{\,\ j} ~.}
The low-energy ${\cal N}=2$ theory is described by a prepotential ${\cal
F}(a^i,\Lambda)$, where the $a^i$ variables, associated to the cycles $A^i$,
are given by the integrals over these cycles of $dS_{SW}$
\eqn\laa{
a^i(u_k,\Lambda) ={1 \over 2\pi i }  \oint_{A^i} {x P_N'(x)\over
\sqrt{P_N^2(x) - 4 \Lambda^{2N}}} ~dx ~.}
The same expression holds for the dual variables $a_{D,i} \equiv \p {\cal
F}/\p a^i$, with $B_{i}$ instead of $A^i$. The effective gauge couplings are
given by \gaugeco. It follows from \defds, \periods\ and \laa\ that 
\eqn\firstders{
{\partial a^i\over \p u_{k+1}}=A^i{_k} ~, \,\,\,\,\,\,\,\,\,\,\,\,\,\
{\partial a_{D,i} \over \p u_{k+1}}=B_{ik} ~.}

The blowup formula arises in the following context. Suppose that we have a
four-manifold $X$, and we consider the blownup manifold at a point $p$, $\hat
X = {\rm Bl}_p(X)$. Under this operation, the homology changes as follows
(see, for example, \gh):
\eqn\homol{
H_2(X) \rightarrow H_2 (\hat X) = H_2 (X) \oplus {\bf Z} \cdot B ~,}
where $B$, the class of the exceptional divisor, satisfies $B^2=-1$. Since the
blownup manifold $\hat X$ has an extra two-homology class, there are extra
operators $I_k(B)$ that must be included in the generating function. We will
then write $\hat S= S+ t B$. There is also the possibility of having a
non-Abelian magnetic flux through the new divisor, and this flux is
specified by a vector $\vec \beta$ with components of the form \mmtwo:
\eqn\flux{
\beta^i =(C^{-1})^i_{\,j}n^j,}
where the $n^j$ are integers, and $(C^{-1})^i_{\, j}$ is the inverse 
of the Cartan matrix for $SU(N)$. The generating function for the 
correlation functions on $\hat X$ is
\eqn\genblow{
\widehat{Z}_{\vec\beta}(p_k,f_k,t_k,\widehat{S})=\biggl\langle \exp\bigl[
\sum_k ( f_k I_k (S) +  t_k I_k(B) + p_k {\cal O}_k) \bigr]
\biggr\rangle_{\widehat{X},\vec\beta} ~,} 
where $t_k =t \cdot \,f_k$. The blowup formula states that this 
generating function is given by 
\eqn\relatione{ 
\widehat{Z}_{\vec\beta}(p_k,f_k,t_k,\widehat{S})=\biggl\langle
\exp\bigl[ \sum_k (f_k I_k (S) + p_k {\cal O}_k )\bigr]  
\tau_{\vec \beta}(t_k|{\cal O}_k) \biggr\rangle_{X} ~,}
where $\tau_{\vec \beta}(t_k|{\cal O}_k)$ will be called the 
{\it blowup function}. This 
function is a series in the $t_k$ whose coefficients are polynomials 
in the operators ${\cal O}_k$:
\eqn\expansion{
\tau_{\vec\beta} (t_k|{\cal O}_k)=\sum_{\vec n} t^{\vec n}{\cal B}_{\vec
n,\vec\beta} ({\cal O}_2, \dots, {\cal O}_N) ~,}
where $\vec n=(n_2, \cdots, n_N)$ is an $(N-1)$-uple of nonnegative integers, 
and $t^{\vec n} \equiv t_2^{n_2} \cdots t^{n_N}_{N}$. The order of the 
terms in the expansion \expansion\ is given by $|\vec n|=\sum_i n_i$. 
The fact that such a formula exists can be justified intuitively by thinking 
about the blowup as a punctual defect which can be represented by an infinite 
series of local operators \mw. Since the ring of local, BRST invariant 
operators is generated by the ${\cal O}_k$, one would expect a factor 
like \expansion\ relating the generating functions. 

The precise expression for the blowup function $\tau_{\vec\beta}(t_k|{\cal
O}_k)$  was derived in \mw\ for the gauge group $SU(2)$, and in \mmtwo\lns\ in
the general case of $SU(N)$, using the $u$-plane integral. To write the
formula for this function, we will need to introduce the Riemann theta
function $\Theta [\vec\alpha, \vec\beta](\vec z|\tau)$ with characteristics
$\vec\alpha=(\alpha_1, \cdots, \alpha_g)$ and $\vec\beta=(\beta_1, \cdots,
\beta_g)$, which we will take as:
\eqn\thetanorm{
\Theta [\vec \alpha, \vec \beta](\vec z|\tau) 
=\sum_{n_i \in \IZ} \exp \Bigl[i\pi \tau_{ij}(n_i+\beta_i)(n_j+ \beta_j) 
+ 2\pi i (n_i+\beta_i)(z_i + \alpha_i)\Bigr] ~.} 
Then, the blowup function has the following form:
\eqn\blowupf{
\tau_{\vec\beta}(t_i|u_i) = {\rm e}^{-\sum_{k,l}t_k t_l \CT_{k,l}}
{\Theta[\vec\Delta,\vec\beta](\vec \xi| \tau) \over \Theta[\vec\Delta,\vec
0](0| \tau)} ~, }
where 
\eqn\vector{
\xi_i= \sum_{k=2}^N {t_k \over 2\pi}{\partial u_{k} \over \partial a^i} ~, 
\,\,\,\,\,\,\,\,\ i=1, \cdots, N-1 ~.} 
We will consider $\vec\beta=\vec 0$ most of the time (notice that, in
general, the $\beta_i$ won't be half-integers). The corresponding blowup
function will be simply denoted by $\tau (t_i| u_i)$. In \blowupf, we have
introduced the symbol $\CT_{k,l}$ to denote the contact terms associated to
the observables $I_k(S)$. They are given  by \lns:
\eqn\confin{
\CT_{k,l}=-{1 \over 2 \pi i }\partial_{\tau_{ij}}
\log \Theta[\vec \Delta,\vec 0] (0|\tau){\partial u_k 
\over \partial a^i}{\partial u_l \over \partial a^j} ~.}
As first noticed in \lns, the explicit expression for the contact terms can
be deduced from the  blowup function by requiring invariance under ${\rm
Sp}(2r, \IZ)$  transformations, and taking also into account that they must
vanish  semiclassically \mw. In the $SU(2)$ case one recovers precisely the
blowup formula of Fintushel and Stern \fs. As remarked in \marijapon, one
of the consequences of the semiclassical  vanishing of the contact terms (or,
equivalently, of the expression \confin) is that the quadratic terms in the
``times'' $t_i$  in the blowup function {\it vanish} for $\vec\beta=\vec 0$,
{\it i.e.} the  expansion \expansion\ has the structure:
\eqn\expaut{
\tau (t_i|u_i)=1 + \sum_{|\vec n|=4} {\cal B}_{(n_2,\cdots,n_N)}(u_i)
~t_2^{n_2} \cdots t_N^{n_N} + \cdots } 
This will be important later on. 
 
\newsec{A survey of the theory of hyperelliptic Kleinian functions}

In the first half of this section we will review in some detail the basic
constructions in the theory of hyperelliptic Kleinian functions. A very good 
modern survey is \victoruno. We will also rely heavily on the results by
Bolza \bolzadif\bolzabrios\bolzarem\ and Baker \baker\bakertwo. In the last
subsection, we will develop a constructive procedure to expand an even
half-integer hyperelliptic $\sigma$-function up to arbitrary order in the
moduli of the curve following the centenarian footsteps of \bolzarem.

\subsec{Hyperelliptic curves and Abelian differentials}

The basic objects we need to develop the theory are Abelian differentials on a
hyperelliptic curve.  Although we will concentrate most of the time on the 
curve \hypersw, we will attempt to give a summary of the general story and
consider hyperelliptic curves of the ``even'' form 
\eqn\hyper{
y^2 = f(x) = \sum_{i=0}^{2g+2}\lambda_i x^i,
} 
describing a Riemann surface of genus $g$. The curve is said to be in {\it
canonical form} when  $\lambda_{2g+2}=0$ and $\lambda_{2g+1}=4$, and any curve
of the form \hyper\ can  be put in such a form by a fractional linear
transformation.  A basis of Abelian differentials of the first kind is given
by the set of $g$ holomorphic 1-forms \holo. To construct hyperelliptic 
$\sigma$-functions, we will also need a basis of Abelian differentials of the
second kind. To introduce these differentials we construct a generating
functional as follows.  First, we consider a function $F(x_1,x_2)$ (sometimes
called a Weierstrass polynomial) which is at most of degree $g+1$ both in $x_1$
and $x_2$, and satisfies the following conditions:
$$
F(x_1,x_2)=F(x_2,x_1) ~,\,\,\,\,\,\ F(x,x)=2f(x) ~,
$$
\eqn\conditions{
\Bigl( {\partial F(x_1,x_2) \over \partial x_1} \Bigr)_{x_1=x_2}= f'(x_2) ~.}
One then defines a basis of Abelian differentials of the second kind, 
$dr^k (x)$ through the identity: 
\eqn\definerre{
\sum_{k=1}^g dv_k(x_1) \, dr^k(x_2) = -{1 \over 2 y_1} {\partial \over
\partial x_2} \Bigl( {y_2 \over x_1-x_2}\Bigr) dx_1\, dx_2 + {F(x_1,x_2)
\over 4(x_1-x_2)^2} {dx_1\, dx_2 \over y_1 y_2} ~,} 
and also a global Abelian differential form of the second kind
\eqn\dospolos{
d\omega (x_1,x_2) = {2 y_1 y_2 + F(x_1,x_2) \over 4(x_1-x_2)^2} ~{dx_1\, dx_2
\over y_1 \, y_2}~,} 
which has a double pole at $x_1=x_2$ with coefficient normalized to $1$. We
will consider three different choices of $F(x_1,x_2)$ in this paper: 

~

\noindent 1) The function used, for example, in \baker\victoruno\ is given by
\eqn\fvictor{
F_{(1)}(x_1,x_2)=2\lambda_{2g+2}x_1^{g+1} x_2^{g+2}+ \sum_{i=0}^g x_1^ix_2^i 
(2\lambda_{2i} + \lambda_{2i + 1}(x_1 + x_2)) ~,}
and the corresponding basis is  
\eqn\basisone{
dr^j =\sum_{k=g+1-j}^{g+j} (k+j-g) ~\lambda_{k-j+g+2} ~{x^k dx \over 4y} ~,}
where $j$ ranges from $1$ to $g$.

~

\noindent 2) A second choice expresses $F(x_1,x_2)$ in a way which is
``covariant'' with  respect to an ${\rm Sl}(2, \IR)$ transformation of the 
$x$-coordinate, as we will explain in more detail in section 5. A convenient 
way to express this polynomial is through the use of a ``symbolic''
notation as follows. The equation for the hyperelliptic curve \hyper\ is
written as 
\eqn\symb{
y^2 = (\alpha_1 + \alpha_2 x)^{2g+2} ~,}
so that 
\eqn\symboltrans{
\lambda_p={2g+2 \choose p} \alpha_1^{2g+2-p} \alpha_2^p ~.}
Of course the notation is symbolic in the sense that $\alpha_1$ and $\alpha_2$
are not defined as complex numbers. One now defines the so-called
$(g+1)$--polar of the hyperelliptic curve as:
\eqn\polar{
F_{(2)}(x_1,x_2) = 2(\alpha_1 + \alpha_2 x_1)^{g+1} (\alpha_1 + \alpha_2
x_2)^{g+1} = 2\sum_{p,q=1}^{g+1} {{g+1 \choose p}{g+1 \choose q} \over {2g+2
\choose p+q}} ~\lambda_{p+q} ~x_1^p x_2^q ~.}
We are not aware of the existence of a simple and closed expression for the
corresponding meromorphic differentials though they can be easily computed
case by case.

~

\noindent 3) A third choice, due to Baker \bakerpaper, and studied in detail 
by Bolza \bolzarem, will be particularly useful in this paper. It is well
known that, for hyperelliptic curves, even and non-singular half-integer
characteristics are in one to one correspondence  with the factorizations of
\hyper\ in two polynomials of degree $g+1$, say $y^2=Q(x)R(x)$. To this
factorization we will associate the Weierstrass  polynomial:
\eqn\bolzapol{
F_{(3)}(x_1,x_2)=Q(x_1) R(x_2) + Q(x_2) R(x_1) ~.}
In the case of Seiberg-Witten hyperelliptic curves \hypersw, the $dr^j$ basis
acquires a simple expression that will be discussed below.

~

It is not difficult to prove that two different choices of Weierstrass 
polynomial, $F(x_1,x_2)$, $\widehat F(x_1,x_2)$, both satisfying 
\conditions, are related by (see \bakertwo, p. 315)
\eqn\relac{
 F(x_1,x_2) - \widehat F(x_1,x_2) = 4(x_1-x_2)^2 \psi (x_1,x_2) ~,}
where $\psi (x_1,x_2)$ is a polynomial symmetric in $x_1,x_2$, and of degree 
at most $g-1$ in each variable. It can therefore be written as 
\eqn\expli{
\psi(x_1,x_2) = \sum_{i,j=1}^g d_{ij} x_1^{g-i} x_2^{g-j} ~,}
where $d_{ij}$ is symmetric in $i,j$. Inserting \relac\ in \definerre\ 
we also obtain the relation between the different basis of second kind 
differentials
\eqn\reldr{
dr^j = \widehat dr^j + \sum_{k=1}^g d_{jk} dv_{k} ~.}

Given a basis of differentials of the second kind $dr^k$, constructed 
from a Weierstrass polynomial $F(x_1,x_2)$, we define the following matrices 
of periods:
\eqn\etaperiods{
\eta^{ki} =-{1\over 2\pi i}\oint_{A^i}dr^k ~,~~~~~~~~~~~~~ \eta'^k_{\,\,\ i}
= -{1\over 2\pi i}\oint_{B_i}dr^k ~.}
These matrices generalize the usual $\eta_{\alpha}=\zeta (\omega_{\alpha})$ 
of an elliptic curve, to a hyperelliptic curve. Notice that the biperiods of
the global Abelian differential \dospolos\ can be written as
\eqn\etaom{
\oint_{A^i} \oint_{A^j} d\omega = 4 \pi^2 ~A^i_{\, k} ~\eta^{kj} ~.}
One can also prove a generalization of Legendre's relation (see, for example,
\victoruno):
\eqn\legendre{ 
\eta = 2 \kappa A ~, ~~~~~~~~~~~~~~~ \eta' = 2 \kappa B - \half (A^{-1})^t ~,}
where $\kappa$ is a symmetric matrix.

\subsec{Hyperelliptic $\sigma$-functions and Kleinian functions}

We are now ready to introduce the key objects: the hyperelliptic 
$\sigma$-functions. To motivate the definition, recall that the usual elliptic 
$\sigma$-functions can be written as quotients of theta functions with an
extra exponential involving the $\eta$-periods (see, for  example, \akhi).
This property  suggests to define the hyperelliptic $\sigma$-functions in
terms  of theta functions. We need to choose a characteristic $[\vec \alpha,
\vec \beta]$ for these  functions, and a Weierstrass polynomial $F(x_1,x_2)$
to define a set of meromorphic Abelian differentials  with their corresponding
$\eta$-periods. The $\sigma$-function is then defined as: 
\eqn\sigm{
\sigma^F[\vec\alpha, \vec\beta](\vec v) = {1\over C} \exp\{v_i \kappa^{il}
v_l\} \Theta[\vec\alpha, \vec\beta ]((2\pi i)^{-1} v_l (A^{-1})^l_{\,\ i}
|\tau) ~.} 
In the above equation, the matrix $\kappa$ (see \legendre) is given by 
\eqn\kap{
\kappa^{il}={1\over 2}\eta^{ij}( A^{-1})^l_{\,\ j} ~,}
and $C$ is a nonzero modular form of weight $(1/2,0)$ with respect to 
the action of ${\rm Sp}(2g, \IZ)$. When the characteristic $[\vec\alpha,
\vec\beta] $ is even and non-singular, a useful choice is the one made in
\bolzadif:
\eqn\constant{ 
C=\Theta[\vec \alpha, \vec \beta](0|\tau)=({\rm det}(A))^{1/2}\Delta_Q^{1/8}
\Delta_R^{1/8} ~,}
where we have used Thomae's formula \fay\ for the even characteristic
associated to the splitting ~$y^2=Q(x) R(x)$, and $\Delta_{Q,R}$ are the 
discriminants of the $Q,R$ factors. 

An important property of the $\sigma$-functions is that they are invariant
under the action of the modular group ${\rm Sp}(2g, \IZ)$. On the other hand,
for a fixed characteristic, $\sigma$-functions corresponding to different
Weierstrass polynomials are related by
\eqn\relatio{
\sigma^{\widehat F}[\vec\alpha, \vec\beta](\vec v) =
\exp{\left(\half \sum_{i,j} d_{ij} v_i v_j\right)} ~\sigma^F[\vec\alpha,
\vec\beta](\vec v) ~,}
where $\sigma^F$ has been defined with $F(x_1,x_2)$ and $\sigma^{\widehat F}$ 
has been defined with $\widehat F(x_1,x_2)$.
 
We are now ready to introduce the hyperelliptic {\it Kleinian functions} as
derivatives of the $\sigma$-function: 
\eqn\kl{
\zeta_j^F[\vec\alpha, \vec\beta ](\vec v) = {\partial \ln \, 
\sigma^F [\vec\alpha, \vec\beta](\vec v) \over \partial v_j} ~, ~~~~~~~~~ 
\wp_{ij}^F[\vec\alpha, \vec\beta](\vec v) = -{\partial^2 \ln \, \sigma^F
[\vec\alpha, \vec\beta ](\vec v) \over\partial  v_i \partial v_j} ~.}
These functions generalize the Weierstrass $\zeta(z)$ and $\wp(z)$ to the
hyperelliptic case, and in some cases they provide an explicit solution for
Jacobi's inversion problem. Notice that they depend, again, on the choice of
Weierstrass polynomial. In particular, one has that 
\eqn\shift{
\wp^{\widehat F}_{ij}[\vec \alpha, \vec \beta](\vec v)= 
\wp^F_{ij}[\vec \alpha, \vec \beta](\vec v) - d_{ij} ~.}
One of the key aspects of the hyperelliptic Kleinian functions
$\wp_{ij}^F[\vec\alpha, \vec\beta](\vec v)$ and of the $\sigma$-functions is
that they satisfy differential equations which generalize those of the
elliptic case like, for example, Weierstrass' cubic relation $(\wp'(u))^2
=4\wp(u)^3-g_2 \wp(u) -g_3$. This will be the subject of the next 
subsection. 
 
\subsec{Differential equations for the hyperelliptic Kleinian functions}
 
The relations involving the hyperelliptic Kleinian functions
$\wp_{ij}^F[\vec\alpha, \vec\beta](\vec v)$ and their derivatives were
originally studied by Baker in \baker\bakertwo. The case of $g=2$ was
investigated in full detail in \baker. A generalization of this construction
has been recently worked out in \victoruno. In this approach, one obtains a
set of second order partial differential equations for the
$\wp_{ij}^F[\vec\alpha, \vec\beta](\vec v)$ with respect to the ``times''
$v_l$, that in principle could be solved in a series expansion. This would
give the series expansion for the $\sigma$-function in terms of the ``times''
and the  moduli of the curve. The main difficulty to extend this method to
higher genus is that the relevant differential equations are given in an
implicit way, and even for $g=3$ a lot of work is needed in order to extract
the first few terms of the expansion (see, for example, \victoruno\ where the
first two terms have been obtained for a special --singular--
characteristic). It is important to notice that these differential
equations, being of second order, are the same for the different
characteristics. The choice of  characteristic shows up in the choice of
initial conditions for the equations.

For the derivatives of $\wp_{11}$ one can, however, write an explicit 
equation for arbitrary genus which will be useful later: 
\eqn\urkdv{
\eqalign{
\wp_{111i}=& (6 \wp_{11} +\lambda_{2g}) \wp_{1i} + {1\over 4} \lambda_{2g+1} 
(6\wp_{i+1,1} -2\wp_{i2} + {1\over 2} \delta_{i1} \lambda_{2g-1}) \cr 
+& {1\over 2} \lambda_{2g+2} (6 \wp_{i+2,1}-6\wp_{i+1,2} + 2\wp_{i3}
-\delta_{i1} \lambda_{2g-2} -{1\over 2} \delta_{i2}\lambda_{2g-3}) ~. \cr}}
The hyperelliptic Kleinian functions which are used in this equation are
defined by means of the Weierstrass polynomial \fvictor. Any other choice
will amount,  by \shift, to a $v$-independent shift. In \urkdv, the extra
subindices denote derivatives with respect  to the components $v_i$. This
equation follows from \victoruno,  eq.(5.3), and when the curve is written
in the canonical way, it reduces to Proposition 4.1 of the same paper.  

~

A different approach to this problem has been taken in a series of papers by
Bolza \bolzadif\bolzabrios\bolzarem, who obtained a partial differential
equation for {\it even} hyperelliptic $\sigma$-functions which can be
explicitly written for any genus. First, Bolza derived an equation for the
logarithmic derivative of the Kleinian functions $\wp^F_{ij}$. Let us
consider a $\sigma$-function defined by the Weierstrass polynomial $F$, and
by the --even and non-singular-- characteristic $[\vec\alpha, \vec\beta]$ 
associated to the factorization $y^2=f(x)=Q(x)R(x)$. Then one has \bolzadif:  
\eqn\bolzath{
\sum_{i,j} \wp_{ij}^F[\vec\alpha, \vec\beta](0)x_1^{g-i} x_2^{g-j} = 
{ F(x_1,x_2)-Q(x_1)R(x_2) - Q(x_2) R(x_1)  \over 4(x_1-x_2)^2} ~.}
This equation will be important later in order to identify the 
$\sigma$-function which is relevant to the blowup formula. Notice, in
particular, that it tells us that $\wp_{ij}^F[\vec\alpha, \vec\beta](0)$
vanishes when the Weierstrass polynomial is $F_{(3)}$.

We can now state Bolza's differential equation for an even $\sigma$-function.
Let $a$ be one of the $2g+2$ zeroes of \hyper. We first define the following 
functions:
\eqn\haches{
(x-z)^{g-1}=\sum_j x^{g-j}h_j(z) ~,} 
and also the matrices $p^F_{ij}(a)$, $q^F_{ij}(a)$ through the relations:
\eqn\matri{
\eqalign{
\sum_{i,j=1}^g p^F_{ij}(a) x^{g-i} h_j (z) = & ~{1\over 2}{(x-z)^{g-1} \over
x-a}-{1\over 2}{(a-z)^{g-1} \over f'(a)} {F(x,a) \over (x-a)^2} ~,\cr
\sum_{i,j=1}^g q^F_{ij}(a) x^{g-i} z^{g-j} = & ~{1\over 8}\Bigl({1 \over x-a}
+ {1 \over z-a}\Bigr){F(x,z) \over (x-z)^2} + {1\over 4}{1 \over
(x-z)^2}{\partial F(x,z)\over \partial a} \cr & ~- {1\over 8} {F(x,a) F(z,a)
\over f'(a)(x-a)^2 (z-a)^2} ~.}} 
In this equation, the $'$ denotes derivative w.r.t. $x$. We can now state the
differential equation satisfied by $\sigma^F[\vec\alpha, \vec\beta](\vec v)$
\bolzadif:
\eqn\bolzadifeq{
{\partial \sigma^F \over \partial a} = - \sum_{i,j=1}^g p^F_{ij}(a) v_i 
{\partial \sigma^F \over \partial v_j} - {1 \over 2} \sigma^F \sum_{i,j=1}^g
q^F_{ij}(a) v_i v_j  + \sum_{i,j=1}^g {a^{2g-i-j} \over f'(a)} \Bigl(
{\partial^2\sigma^F \over \partial v_i \partial v_j} + \sigma^F
\wp^F_{ij}(0) \Bigr) ~,}
where we have dropped the characteristic to gain in clarity. This equation
endowes recursive relations for the Taylor expansion of $\sigma^F$. In fact,
the appearance of a set of recursive relations is immediate provided we
replace our even $\sigma$-function by its Taylor expansion
\eqn\taylorexp{
\sigma^F[\vec\alpha, \vec\beta](\vec v) = \sum_{n=0}^\infty {1 \over (2n)!}
~\varsigma_n(\vec v) ~,}
where $\varsigma_n(\vec v)$ are homogeneous polynomials of degree $2n$ in
$v_l$,
\eqn\homog{
\sum_{i=1}^g v_i {\partial \varsigma_n(\vec v) \over \partial v_i} =
2n ~\varsigma_n(\vec v) ~. }
The recursive relation for the $\varsigma_n$ polynomials reads
\eqn\recursive{
\eqalign{
\sum_{i,j=1}^g {a^{2g-i-j} \over f'(a)} & {\partial^2\varsigma_n \over
\partial v_i \partial v_j} = 2n (2n-1) \Big\{{\partial \varsigma_{n-1} \over
\partial a} - \varsigma_{n-1} \sum_{i,j=1}^g {a^{2g-i-j} \over f'(a)}
~\wp^F_{ij}(0) \cr & + \sum_{i,j=1}^g p^F_{ij}(a) v_i {\partial
\varsigma_{n-1} \over \partial v_j} + (n-1) (2n-3) ~\varsigma_{n-2}
\sum_{i,j=1}^g q^F_{ij}(a) v_i v_j \Big\} ~. }}
The main difficulty of these equations is that they involve the derivatives of
$\sigma^F$ or $\varsigma_n$ with respect to a branch point $a$, which is of
little practical use. However, as we will see in what follows, one can
deduce from \bolzadifeq\ a differential equation involving the coefficients 
of the curve \hyper, which will allow us to give recursive relations for the
expansion of the $\sigma$-functions relevant to the blowup formula. A final
comment is in order. Due to the fact that $\varsigma_0 = 1$, one can already
obtain a set of differential equations for the quadratic term from \recursive\
that yields
\eqn\cuadratic{
\varsigma_1 (\vec v) = - \sum_{i,j=1}^g \wp^F_{ij}(0) v_i v_j ~.}
Thus, after \bolzath, the quadratic contribution to the $\sigma$-function
vanishes when $F = F_{(3)}$.

Let us fix for future convenience the Weierstrass polynomial to be $F_{(3)}$
\bolzapol . This implies no lack of generality as long as \relatio\ 
allows to go from a given polynomial to any other. The corresponding matrices
$p^{(3)}_{ij}(a)$ and $q^{(3)}_{ij}(a)$ get further simplified to:
\eqn\matrif{
\eqalign{
\sum_{i,j=1}^g & p^{(3)}_{ij}(a) x^{g-i} h_j (z) = {(x-z)^{g-1} \Xi(a,a) -
(a-z)^{g-1} \Xi(x,a) \over 2 ~\Xi(a,a) (x-a)} ~, \cr
\sum_{i,j=1}^g & q^{(3)}_{ij}(a) x^{g-i} z^{g-j} = {\Xi(x,z) \Xi(a,a) -
\Xi(x,a) \Xi(z,a) \over 8 ~\Xi(a,a) (x-a) (z-a)} ~,}}
where we have introduced the quantity $\Xi(x,z)$,
\eqn\bolzaph{
\Xi(x,z) = {Q(x) R(z) - Q(z) R(x) \over x - z} ~,}
that can be easily seen to be a symmetric polynomial of degree at most $g$ in
its variables. We shall assume in what follows that $a$ is a root of $Q(x)$.
Notice then that $\Xi(x,a) = Q(x) R(a)/(x-a) = - R(a) \partial Q(x)/\partial
a$ and $\Xi(a,a) = f'(a) = Q'(a) R(a)$. We also have that $\wp^{(3)}_{ij}(0)$
vanishes. The recursive equation \recursive\ can be written as
\eqn\recursivetwo{
\eqalign{
\sum_{i,j=1}^g a^{2g-i-j} {\partial^2\varsigma_n \over \partial v_i \partial
v_j} & = 2n (2n-1) ~\Xi(a,a)~ \Big\{{\partial \varsigma_{n-1} \over \partial
a} + \sum_{i,j=1}^g p^{(3)}_{ij}(a) v_i {\partial \varsigma_{n-1} \over
\partial v_j} \cr & + (n-1) (2n-3) ~\varsigma_{n-2} \sum_{i,j=1}^g
q^{(3)}_{ij}(a) v_i v_j \Big\} ~.}}
Now, let $\varphi(x)$ be a polynomial of degree $g+p$. Then, one has
\eqn\lemma{
\sum_{(a)} {\Xi(x,a) \over \Xi(a,a)} ~\varphi(a) = \varphi(x) -
\sum_{i=0}^{p-1} \mu_i x^i Q(x) ~,} 
for appropriate $\mu_i$ defined in such a way that the polynomial of the r.h.s.
has degree $g$. This result comes immediately from the fact that both sides of
\lemma\ are equal when evaluated at any of the $g+1$ roots of $Q(x)$. Consider
now the function
\eqn\eme{
M(x,z) = \Xi(x,z) Q'(z) - \sum_{i=0}^{g-1} \mu_i(x) z^i Q(z) ~,}
where, again, $\mu_i(x)$ are chosen in such a way that $M(x,z)$ is of degree
$g$ in variable $z$,
\eqn\expeme{
M(x,z) = \sum_{i=1}^{g+1} m_i(x) z^{g+1-i} ~.}
It can be written, after \lemma, as
\eqn\lemmatwo{
M(x,z) = \sum_{(a)} {\Xi(z,a) \over \Xi(a,a)} ~M(x,a) = - \sum_{(a)} \Xi(x,a)
{\partial Q(z) \over \partial a} = - \sum_{(a)} \Xi(x,a) \sum_{i=0}^{g+1}
{\partial q_i
\over \partial a} z^{g+1-i} ~,}
where $q_i$ are the coefficients of $Q(x)$. For a given function ${\cal G}$,
we can replace $z^{g+1-i}$ by $\partial{\cal G}/\partial q_i$ in
\expeme\lemmatwo\ with the result
\eqn\bothderiv{
\sum_{i=1}^{g+1} m_i(x) {\partial{\cal G} \over \partial q_i} = - \sum_{(a)}
\Xi(x,a) {\partial{\cal G} \over \partial a} ~.}
We now multiply \recursivetwo\ by $\Xi(x,a)/\Xi(a,a)$ and sum over $(a)$. The
l.h.s. as well as the last two terms of the r.h.s. are poynomials in $a$,
while the remaining term is the above referred problematic derivative that we
can now handle by means of \bothderiv. Conversely, we can instead consider a
root $b$ of the polynomial $R(x)$ and arrive to formulae analogous to
\lemma--\bothderiv\ with $b, R$ and $r_i$ instead of $a, Q$ and $q_i$,
whereas $m_i(x)$ is replaced by, say, $-n_i(x)$ due to the change of sign of
$\Xi(x,b)$ with respect to $\Xi(x,a)$. At the end of the day, the recursive
relation can be brought to the following form
\eqn\recursivethree{
{\cal Z}[x,\varsigma_n(\vec v)] = - 2n (2n-1) \Big\{ \Delta \varsigma_{n-1} -
{\cal P}[x,\varsigma_{n-1}(\vec v)] - (n-1) (2n-3) ~\varsigma_{n-2}~ {\cal
Q}[x,\vec v] \Big\} ~, }
where the polynomials ${\cal Z}[x,\varsigma_n(\vec v)]$, ${\cal
P}[x,\varsigma_{n-1}(\vec v)]$ and ${\cal Q}[x,\vec v]$,
\eqn\inshort{
\eqalign{
{\cal Z}[x,\varsigma_n(\vec v)] & \equiv \sum_{(a)} {\Xi(x,a) \over \Xi(a,a)}
~\sum_{i,j=1}^g a^{2g-i-j} {\partial^2\varsigma_n \over \partial v_i \partial
v_j} + (a \to b) ~, \cr 
{\cal P}[x,\varsigma_{n-1}(\vec v)] & \equiv \sum_{(a)} \Xi(x,a)
\sum_{i,j=1}^g p_{ij}(a) v_i {\partial \varsigma_{n-1} \over \partial v_j} -
(a \to b) ~, \cr  {\cal Q}[x,\vec v] & \equiv \sum_{(a)} \Xi(x,a)
\sum_{i,j=1}^g q_{ij}(a) v_i v_j + (a \to b) ~,}}
should be computed as explained in \lemma, and the differential operator
$\Delta$ is given by
\eqn\diffop{
\Delta \varsigma_{n-1} = \sum_{i=1}^{g+1} m_i(x) {\partial \varsigma_{n-1}
\over \partial q_i} + \sum_{i=1}^{g+1} n_i(x) {\partial \varsigma_{n-1} \over
\partial r_i} ~.}
Notice that $\Delta$ involves derivatives with respect to all the coefficients
of the hyperelliptic curve. Thus, when considering the setup provided by the
Seiberg-Witten geometry, it will be necessary to retain the dependence of any
quantity on the whole set of coefficients of the curve, provided one is
interested in higher orders of the Taylor expansion. The procedure described
above leads to a recursive computation of the hyperelliptic $\sigma$-function
up to arbitrary order in time variables.

\newsec{Expansion of the blowup function}

We will show in this section that the formalism discussed above is the
appropriate framework to address a detailed study of the blowup function.
 
\subsec{The Seiberg-Witten geometry}
 
We will be now more specific and focus on the hyperelliptic curve \hypersw\ 
describing the low-energy effective action of ${\cal N}=2$, $SU(N)$ 
super Yang-Mills theory. This curve can be written as follows: 
\eqn\factor{
y^2(x)=Q(x)R(x) ~,}
where
\eqn\factores{
Q(x)=P_{N}(x)-2\Lambda^N ~,\,\,\,\,\,\,\,\,\ R(x)=P_{N}(x)+2\Lambda^N ~.}
The Weierstrass polynomial which is relevant to our problem is, as will be
clear below, $F_{(3)}$ \bolzapol. It is not difficult to prove  that the
Abelian differentials of the second kind corresponding to this  generating
function are given by:
\eqn\drzero{
dr^j={1\over 2}P'_{j}(x) P_{N}(x){dx \over y} ~, ~~~~~~~ j=1, \cdots, N-1 ~.}
From now on, unless the contrary is stated, the $dr^j$ will denote the above
differentials, {\it i.e.}  we will assume that the basis of Abelian
differentials is given by the  generating function \bolzapol\ for the specific
case of the Seiberg-Witten  curve \hypersw. Notice that 
\eqn\drone{
dr^1 = {1 \over 2N} dS_{SW} -{1\over 2N} \sum_{k=1}^g (k+1)u_{k+1}dv^k ~.}
These Abelian differentials of the second kind are associated to the 
coordinates on the Jacobian $v_i$, $i=1, \cdots, N-1$, that appear in the 
expression for the $\sigma$-function \sigm. In this sense, they play the 
role of the differentials that define a Whitham hierarchy and a prepotential 
theory \Krichever. 

The connection to the Whitham hierarchy can be made more concrete by relating
the differentials \drzero\ to another basis of Abelian differentials of the 
second kind which will be useful later. This basis was introduced in \ITEP,
and is given by 
\eqn\sombreros{
d\hat \Omega_n =R_{n,N}(x){P_{N}'(x) dx\over y} ~,}
where the polynomials $R_{n,N}(x)$, of degree $n$, are given by $R_{n,N}(x) =
(P_{N}(x))_+^{n\over N}$.  In this equation, $(P_{N}(x))^{n\over N}$ denotes
the $n/N$-th power of the polynomial $P_{N}(x)$ understood as a Laurent
series in $x$
\eqn\bnm{P_{N}(x)^{n\over N}=\sum_{m=-\infty}^n b_{m,n}x^n ~,} 
and the $+$ suffix means that one only keeps the nonnegative powers of $x$.
One has, for example \ITEP:
\eqn\primpol{
\eqalign{
&R_{1,N}(x) = x ~, \,\,\,\,\,\,\,\,\ R_{2,N}(x)= x^2 -{2 \over N}u_2 ~,\cr
&R_{3,N}(x) = x^3 - {3 \over N}u_2 x  -{3\over N}u_3 ~.\cr}}
So, in particular, $d\hat \Omega_1 = dS_{SW}$. The relation between these
polynomials and \drzero\ is:
\eqn\relacion{
d\hat{\Omega}_n = {2N\over n}\sum_{p=1}^n b_{n-N,p-N} ~dr^{N-p} -
\sum_{m=1}^{N-1}a_{n,m} dv_m ~,}
where 
\eqn\anms{
a_{n,m} = \sum_{p=0}^{N-m-1}(N-m-p) \left(~b_{n,p}~u_{m+p} + {N \over n}
\sum_{k=1}^{n} b_{n-N,-k}~u_{m+p-k}~u_{N-p}\right) ~.} 
taking $u_0=-1$, $u_1=0$ and $u_{k>N} = u_{k<0} = 0$. For $N=3$, for example,
one finds:
\eqn\gtwoex{
\eqalign{
d\hat \Omega_1= & 6 dr_1 + 2 u_2 dv_1 + 3u_3 dv_2 ~,\cr
d\hat \Omega_2 =& 3 dr_2 + 3 u_3 dv_1 + {2 u_2^2\over 3} dv_2 ~.\cr}}
It is precisely the basis $d\hat{\Omega}_n$ the one that turns out to be
relevant in the study of adiabatic deformations of the Seiberg-Witten
solution within the framework of the Whitham hierarchy \ITEP.

\subsec{Blowup function and $\sigma$-functions. Contact terms revisited}

We will only consider in this section the case of zero magnetic flux, so
$\vec\beta=\vec 0$ and the characteristic of the theta function is
$[\vec\Delta,\vec 0]$. This characteristic is the one associated to the
splitting  of the Seiberg-Witten curve given in \factor\ (see \ITEP\emm).  In
view of \sigm, we see that the blowup function \blowupf\ has the form  of a
$\sigma$-function. To make this comparison more precise, notice that, in the
Seiberg-Witten context, 
\eqn\identif{
(A^{-1})^l_{\,\ i} = {\partial u_{l+1} \over \partial a^i} ~, ~~~~~~~~~
\kappa^{il}= {1\over 2}\eta^{ij}{\partial u_{l+1} \over \partial a^j} ~.}
This means that the ``times'' of the blowup function are related to the vector
$\vec v$ in \sigm\ just by $v_l=it_{l+1}$. We have to compare now the
exponentials in \sigm\ and \blowupf.  As we stressed at the end of section 2,
when there is no non-Abelian magnetic flux through the exceptional divisor,
{\it i.e.} the characteristic is $[\vec\Delta,\vec 0]$, the quadratic terms in
the expansion of the blowup function vanish \expaut. But this is precisely the
behavior of the $\sigma$-function associated to the generating function
\bolzapol, as  it follows from \bolzath\ and \cuadratic. We then obtain the
following results: 

~

\noindent
$\bullet$ The blowup function of $SU(N)$ Donaldson theory in the absence of
magnetic flux is a hyperelliptic $\sigma$-function with characteristic
$[\vec\Delta,\vec 0]$ and with the Weierstrass polynomial given in \bolzapol,
\eqn\taus{
\tau(t_i|u_i)=\sigma^{F_{(3)}}[\vec\Delta,\vec 0](it_{l+1}) ~.}
This identity, combined with the results of section 3, gives a rather explicit
realization of the expansion \expansion. We will give concrete results for the
lower genus hyperelliptic surfaces in the next subsection.

~

\noindent
$\bullet$ The contact terms $\CT_{k+1,l+1}$ are given by 
\eqn\cont{
\CT_{k+1,l+1} = \kappa^{k,l} = - {1\over 8 \pi i}{\partial u_{l+1} 
\over \partial a^i}\oint_{A^i}P_k'(x)P_N(x){dx \over y} ~,}
where $\kappa$ is the matrix introduced in \legendre, and we have used 
the explicit expression for the $dr^k$ given in \drzero. This result gives
yet another remarkably simple form of writing the contact terms of $SU(N)$
twisted Yang-Mills theory, this time in terms of periods of Abelian
differentials. Using \drone, one obtains, for example:
\eqn\contu{
\CT_{2,\ell}={1 \over 4N}\biggl( \ell u_{\ell}- a^i{\partial u_{\ell} \over
\partial a^i} \biggr) ~,} 
for $l=1, \cdots, N-1$. One can in fact explicitly check some of these
expressions by using the results of \ITEP\emm. The starting point are the
Whitham equations
\eqn\whith{
\Bigl( {\partial u_k \over \partial \log \Lambda} \Bigr)_
{ T_{n \ge 2}=0} = k u_k - a^i {\partial u_k \over \partial 
a^i} ~,\,\,\,\,\,\,\,\,\,\,\,\,\,\  
\Bigl( {\partial u_k \over \partial  T_n} 
\Bigr)_{T_{n \ge 2}=0} = - c_{(n)}^i {\partial u_k \over \partial 
a^i} ~,}
where, in the second equation, $n=2, \cdots, N$, and 
\eqn\ces{
c_{(n)}^i= {1 \over 2 \pi i} \oint_{A^i} d\widehat \Omega_n ~.}
In \whith, the slow times $T_n$ are the ``hatted times'' introduced in \emm.
The Whitham equations in the above form can be easily deduced from equation
(3.18) of \ITEP\ and the redefinition of Whitham times in \emm. Notice that
these equations have already the flavor of \cont, since they express the
derivatives of the moduli with respect to the slow times in terms of
$A$-periods of Abelian differentials of the second kind. The derivatives of
the moduli entering in \whith\ are in fact closely  related to the contact
terms. In the formalism of \ITEP, the natural  duality-invariant coordinates
are not the moduli $u_{k+1}$, but some combinations thereof: 
\eqn\modulorus{
{\cal H}_{k+1, l+1}= {N \over kl} {\rm res}_{\infty}\Bigl[ 
(P_{N}(x))^{k\over N}d(P_{N}(x))_+^{l\over N} \Bigr] ~.}
The moduli $u_{k+1}$ are substituted in this formalism by:
\eqn\casidef{
{\cal H}_{k+1} \equiv {\cal H}_{k+1,2} = u_{k+1} +
g_{k+1}(u_2,\cdots,u_{k-1}) ~. }
One has, for example: 
\eqn\exa{
{\cal H}_2 = u_2 ~, \,\,\,\,\,\,\,\,\ {\cal H}_3 = u_3 ~, \,\,\,\,\,\,\,\,\
{\cal H}_{3,3} = u_4 + {N-2\over 2N} ~u_2^2 ~.}
The RG equations of \ITEP\ give explicit results for the derivatives 
of the ${\cal H}_{k+1}$:
\eqn\secders{
\eqalign{
\Bigl({\partial {\cal H}_{k+1} \over \partial \log \Lambda}\Bigr)_{T_{n\ge
2}=0} =&-2N {\partial {\cal H}_2 \over \partial a^i} {\partial {\cal H}_{k+1}
\over \partial a^j}{1 \over \pi i} \partial_{\tau_{ij}}\log \Theta[\vec\Delta,
\vec 0](0|\tau) ~, \cr
\Bigl( {\partial {\cal H}_{k+1} \over \partial T_{l}} \Bigr)_{T_{n\ge 2}=0} =&
-(k+l){\cal H}_{k+l}-{2N\over l} {\partial {\cal H}_{k+1} \over \partial a^i}
{\partial {\cal H}_{l+1} \over \partial a^j}{1 \over \pi i}
\partial_{\tau_{ij}}\log \Theta[\vec \Delta, \vec 0](0|\tau) ~. \cr}} 
Since the first equation in \secders\ also holds by substituting 
${\cal H}_{k+1} \rightarrow u_{k+1}$, one can combine it with \confin\ and 
\whith\ to obtain precisely \contu. In the same way, one can obtain
expressions relating the $\CT_{k,l}$ to the periods of the family of Abelian
differentials \sombreros, and then use \relacion\ to check \cont. For
example, for $g=2$ one finds: 
\eqn\conts{
\CT_{3,3} = {u_2^2 \over 9} - {1\over 12} {\partial u_3 \over 
\partial a^i} ~c^i_{(2)} ~.}
Using now \gtwoex\ one can explicitly check \cont\ for $SU(3)$ twisted
Yang-Mills theory. 

The expression \cont\ for the contact terms turns out to be very useful, since
the differentials $dr^j$ are rather explicit  in comparison with the Abelian
differentials $d\hat \Omega_n$  introduced in \ITEP. In particular, there are
some cases in which \cont\ is more effective than the expression \confin\ 
involving theta functions. We will see an example in section 6. There, we
treat in detail the case of manifolds of simple type, where contributions
come only from those points of the moduli space where the maximal number of
mutually local monopoles (dyons) get simultaneously massless. 
   
\subsec{Expansion of the blowup function for lower genus}

It is instructive to consider in more detail the way in which the formalism of
the previous section leads to an expansion of the blowup function as in
\expansion\ for hyperelliptic surfaces of lower genus. We already know the
answer for the first two terms, since $\tau(t_i|u_i)$ is an even
$\sigma$-function with generating function $F^{(3)}$: $\varsigma_0 = 1$ and
$\varsigma_1 = 0$. The differential equations for the fourth order term are
encoded in the relation
\eqn\diffcuar{
{\cal Z}_{SW}[x,\varsigma_2(\vec v)] = 12 ~{\cal Q}_{SW}[x,\vec v] ~,}
where we use the subindex $SW$, to indicate that a given quantity has been
evaluated in the Seiberg-Witten curve \factor. The l.h.s. of \diffcuar\ is
given by
\eqn\lowquz{
{\cal Z}_{SW}[x,\varsigma_2(\vec v)] = 2 \sum_{i,j=1}^g x^{2g-i-j}
\varsigma^{(ij)}_2  - \sum_{i=0}^{g-3} (\mu_i Q(x) + \nu_i R(x)) x^i ~,}
where $\mu_i$ and $\nu_i$ are constants (with respect to $x$ though functions
of the ``times'' $\vec v$) that reduce the degree of \lowquz\ as explained in
\lemma, and we have defined
\eqn\defi{
\varsigma^{(ij)}_n \equiv {\partial^2\varsigma_n \over \partial v_i \partial
v_j} ~.}
Notice that the second term of the r.h.s. in \lowquz\ vanishes for $g=2$. We
obtained, for example,
\eqn\zetados{
\eqalign{
{\cal Z}_{SW}[x,\varsigma_2(\vec v)] = & ~2 \Bigl[ \varsigma^{(11)}_2
 ~x^2 + 2 \varsigma^{(12)}_2 ~x + \varsigma^{(22)}_2 \Bigr] ~, \cr
{\cal Z}_{SW}[x,\varsigma_2(\vec v)] = & ~2 \Bigl[ 2 \varsigma^{(12)}_2 ~x^3 +
(\varsigma^{(22)}_2 + 2 \varsigma^{(13)}_2 + u_2 \varsigma^{(11)}_2) ~x^2 +
(2 \varsigma^{(23)}_2 + u_3 \varsigma^{(11)}_2) ~x \cr
& + (\varsigma^{(33)}_2 + u_4 \varsigma^{(11)}_2) \Bigr] ~, \cr}}
for $g=2$ and $g=3$ respectively. Concerning ${\cal Q}_{SW}[x,\vec v]$, a
closed expression does not seem to be feasible. In the cases of lower genus,
we found 
\eqn\qdos{
\eqalign{
{\cal Q}_{SW}[x,\vec v] = & ~-4\Lambda^6 (v_2^2 ~x^2 + 4 v_1 v_2 ~x +
v_1^2 + u_2 v_2^2) ~, \cr
{\cal Q}_{SW}[x,\vec v] = & ~-4\Lambda^8 (4 v_2 v_3 ~x^3 + (6 v_1 v_3 + 4
v_2^2 - u_2 v_3^2) ~x^2 + (4 v_1 v_2 + 3 u_3 v_3^2) ~x \cr
& + (u_4 + u_2^2) v_3^2 + v_1^2 - 2 v_3 (u_2 v_1 - u_3 v_2)) ~. \cr}}
Inserting these polynomials in \diffcuar\ results in a set of differential
equations for $\varsigma_2$ that can be easily solved. For example, in the
case of $g=3$, {\it i.e.} $N=4$, the resulting expansion for the blowup
function is
\eqn\sigtres{
\tau_{SU(4)}(t_i|u_i) = 1 - {\Lambda^8\over 12} \Bigl[ u_2^2 t_4^4 - 4 u_2
t_4^3 t_2 + 4 u_3 t_4^3 t_3 + 6 t_2^2 t_4^2 + 12 t_2 t_3^2 t_4 + 2 t_3^4
\Bigr] + \cdots }
In the case $g=2$, {\it i.e.} $N=3$ it is interesting to work out in detail
the next-to-leading order in the expansion. Notice that it is only
 from the sixth order
term in the Taylor expansion of the blowup function that the full complexity
of \recursivethree\ enters into the game. Thus, for the sake of checking the
recursive procedure that we derived in the previous section we must compute
$\varsigma_3$. The relevant equation is
\eqn\diffsix{
{\cal Z}_{SW}[x,\varsigma_3(\vec v)] = - 30 \Big\{ \Delta \varsigma_{2}|_{SW}
- {\cal P}_{SW}[x,\varsigma_{2}(\vec v)] \Big\} ~, }
where we must include the full dependence of $\varsigma_{2}$ in the
coefficients of a generic hyperelliptic curve before applying the differential
operator $\Delta$. The second term of the r.h.s. in \diffsix\ vanishes for
$g=2$. On the other hand, the term in the l.h.s. is exactly as \zetados\
provided we replace $\varsigma_2$ by $\varsigma_3$. The final answer for the
blowup function up to sixth order in the ``times'' is
\eqn\sigdos{
\eqalign{
\tau_{SU(3)}(t_i|u_i) = & ~1 - {\Lambda^6\over 12} \Bigl[ u_2 t_3^4 + 6
t_2^2 t_3^2 \Bigr] - {\Lambda^6\over 360} \Big[ 3 t_2^6 - 15 u_2 t_2^4 t_3^2 -
60 u_3 t_2^3 t_3^3 - 15 u_2^2 t_2^2 t_3^4 \cr
& - 12 u_2 u_3 t_2 t_3^5 - u_2^3 t_3^6 + 3 u_3^2 t_3^6 - 12 \Lambda^{6} t_3^6
\Bigr] + \cdots }}
Notice that $\tau(t_i|u_i)$ is homogeneous of degree zero provided we assign a
negative weight $1-i$ to variables $t_i$. We will use the expansions \sigtres\
and \sigdos\ in section 6  below to check the expressions for the blowup
functions in the  case of manifolds of simple type. 

\newsec{Relation with the KdV hierarchy}

In this section, we will show that the blowup function satisfies the
differential equations of the KdV hierarchy. More precisely, we will show
that, after  redefining the times through a linear transformation, we obtain a 
$g$-gap solution of the KdV hierarchy. This is essentially a consequence 
of Theorem 4.6 of \victoruno\ (which we review below), but some extra work is
needed in order to adapt it to our context. We first analyze the  effect of
special linear transformations on the hyperelliptic $\sigma$-functions, and
then we establish the relation with the KdV hierarchy. A similar relation has
been pointed out in \Tak. 

\subsec{${\rm Sl}(2, \IR)$ covariance of the $\sigma$-functions}

Consider a hyperelliptic curve of degree $2g+2$ written in the symbolic 
form \symb, and perform an ${\rm Sl}(2, \IR)$ transformation of the 
$x$-variable: 
\eqn\lineart{
x={a + bt \over c + dt} ~, \,\,\,\,\,\,\,\ bc-ad =1 ~.}
The curve \symb\ becomes
\eqn\bec{
Y^2 = (\beta_1 + \beta_2 t)^{2g+2} = \sum_{i=0}^{2g+2} \widehat \lambda_i t^i
~,} 
where 
\eqn\newvar{
Y=(c + dt)^{g+1}y ~, \,\,\,\,\,\,  \beta_1 =c \alpha_1 + a \alpha_2 ~, 
\,\,\,\,\,\,\ \beta_2 =d \alpha_1 + b \alpha_2 ~.}
It is clear that one can always choose the ${\rm Sl}(2, \IR)$ transformation 
in such a way that the new curve is in canonical form, {\it i.e.} such that  
\eqn\gocan{
\widehat \lambda_{2g+2}=\beta_2^{2g+2}=0, \,\,\,\,\,\,\,\,\ 
\widehat \lambda_{2g+1}=\beta_1 \beta_2^{2g+1}=4 ~.}

We will now analyze the changes induced by this transformation in the rest of
the objects defining the $\sigma$-functions. First, we consider the Abelian
differentials of the first kind \holo. Since 
\eqn\relholo{
x^{g-i} {dx \over y} = (a + bt)^{g-i} (c+ dt)^{i-1} {dt \over Y} ~,} 
for $i=1, \cdots, g$, it follows that
\eqn\rel{
dv_i(x) = \Lambda_i^{\,m} d\hat v_m (t) ~,}
where the matrix $\Lambda_i^{\, m}$ can be obtained from the ${\rm Sl}(2,\IR)$ 
matrix by using \relholo. This matrix is invertible, since one can explicitly
construct an inverse by writing $t=(cx-a)/(b-dx)$. It follows  from \relholo\
that the periods of the Abelian differentials of the  first kind transform as:
\eqn\transper{
A^i_{\,j}=\widehat A^i_{\,m}\Lambda_j^{\, m} ~, \,\,\,\,\,\,\,\,\ 
B_{ij}=\widehat B_{im}\Lambda_j^{\,m} ~,}
and therefore the period matrix $\tau$ remains invariant under this 
transformation (in the above equations, the hat refers to the periods of 
the curve \bec.) 

Let us now examine the $\eta$-periods. We have to make now a choice of
Weierstrass polynomial, and to achieve covariance under ${\rm Sl}(2,\IR)$ we
take \polar. It is easy to check that 
\eqn\ftrans{
F_{(2)}(x_1,x_2) = (c+dt_1)^{-g-1} (c+dt_2)^{-g-1} F_{(2)}(t_1,t_2) ~.} 
Therefore, the normalized global Abelian differential of the second kind
\dospolos\ also remains invariant. From \etaom, one can then deduce the
transformation properties of the $\eta$-periods:
\eqn\etatrans{
\widehat \eta ^{ij} = \Lambda_k^{\,\, i} \eta^{kj} ~.}
We can now examine the properties of the $\sigma$-function under these 
transformations. Define 
\eqn\newtimes{
\widehat v_l =(\Lambda^{-1})_l^{\, m} v_m ~,}
which is nothing but a linear transformation of the ``evolution times''. 
Using the above results, we find that
\eqn\sigtrans{
\sigma^F [\vec\alpha, \vec\beta](v_l)_{(x,y)} =
\sigma^F [\vec\alpha, \vec\beta](\widehat v_l)_{(t,Y)} ~,}
where $F$ denotes here the polar Weierstrass polynomials associated to the
corresponding curves. This is the key result that we will need. An important 
corollary of \sigtrans\ is that, after substituting $v_l =\Lambda_l^{\, m}
\widehat v_m$, the $\sigma$-function $\sigma^F [\vec\alpha,
\vec\beta](v_l)_{(x,y)}$ satisfies the same differential  equations than 
$\sigma^F [\vec \alpha, \vec \beta ](\widehat v_l)_{(t,Y)}$ with respect 
to the hatted times. 

\subsec{The KdV hierarchy}

One of the key results of \victoruno\ is that the hyperelliptic Kleinian
functions satisfy the equations of the KdV hierarchy, when the curve is
written in a canonical form, and when the Weierstrass polynomial is given by
\fvictor. This can  be easily deduced from \urkdv. When $\widehat
\lambda_{2g+2}=0$, $\widehat \lambda_{2g+1}=4$, the equation becomes: 
\eqn\ursukdv{
\wp_{111i} =(6\wp_{11}+ \widehat \lambda_{2g})\wp_{1i} + 6\wp_{i+1,1}
-2\wp_{2i} + {1\over 2} \delta_{i1} \widehat \lambda_{2g-1} ~.}
Take now ${\cal U}=2 \wp_{11} + {1\over 6} \widehat \lambda_{2g}$, put 
$x\equiv v_1$ and let $t_i=v_i$ be the higher evolution times. The equation
\ursukdv\ reads:
\eqn\kdveqn{
{\partial {\cal U} \over \partial t_2}={1 \over 4}{\cal U}'''-{3\over 2} 
{\cal U} {\cal U}' ~,} 
where $'$ denotes derivatives w.r.t. $x$. \kdveqn\ is precisely the KdV
equation. It is easy to prove that in fact ${\cal U}$ solves the KdV
hierarchy, or, more precisely, that it is a $g$-gap solution of the
hierarchy. To see this, recall that the higher  evolution equations of the
KdV hierarchy  are (for a review, see Appendix A of \zj):
\eqn\kdvhier{
{\partial {\cal U} \over \partial t_i}=R_i'({\cal U}, {\cal U}', \cdots) ~, 
\,\,\,\,\,\,\,\,\ i\ge 3 ~,}
where the functions in the right hand side are defined recursively as follows:
\eqn\recur{
R_{i+1}'={1 \over 4} R_i'''- ({\cal U}+c)R_i'-{1\over 2}{\cal U}'R_i ~,}
and $c$ is a constant. The equations \kdvhier\ and \recur\ with $c=\widehat
\lambda_{2g}/12$  can be easily checked using again \ursukdv\ and 
\eqn\another{
\wp_{111}\wp_{1i}-\wp_{11i}\wp_{11}+ \wp_{11,i+1}-\wp_{12i}=0 ~, }
which is obtained from \ursukdv\ by imposing $\partial_i\wp_{1111} =
\partial_1 \wp_{111i}$. 

We can now state our main result about the relation of the blowup function to
the KdV hierarchy. Taking into account \taus\ and \sigtrans, we can
write: 
\eqn\carakdv{
\tau(v_m=\Lambda_m^{\,\,\ l}\widehat v_l|{\cal O}_i) =
{\rm e}^{\sum_{ij}c_{ij}\widehat v_i \widehat v_j} \sigma^F[\vec\Delta, \vec
0](\widehat v_l)_{(t,Y)} ~,} 
where the $\sigma$-function in the right hand side has been defined using the
Weierstrass function \fvictor, and the linear transformation $\Lambda$ has
been chosen in such a way that the  hyperelliptic curve $(t,Y)$ is written in
a canonical form. The $c_{ij}$  are constants depending on the parameters of
the ${\rm Sl}(2, {\IR})$  transformation and the moduli of the curve, and 
they can be computed explicitly. They simply arise as in \relatio, by 
comparing $\sigma$-functions defined for different Weierstrass polynomials. 
Using the results above, we finally find that  
\eqn\kdvfin{
{\cal U} = -2 {\partial^2 \log\tau \over \partial \widehat v_1^2} + 4 c_{11}
+ {1 \over  6} \widehat \lambda_{2g}}
is a $g$-gap solution of the KdV hierarchy. In other words, the blowup 
function is, up to a redefinition of the evolution times and the shift in
\kdvfin,  a $\tau$-function of the KdV hierarchy. Remember that the blowup
function  appears in fact in the generating function of the correlation
functions  involving the exceptional divisor. A corollary of the above is that 
these correlation functions on the manifold $\widehat X$ are governed 
by the KdV hierarchy, and they have as initial conditions the generating 
function of the original manifold $X$. 

In \fs, the blowup function of $SU(2)$ Donaldson-Witten theory was obtained 
precisely by solving a differential equation. The above result shows that 
the generalization to $SU(N)$ involves the KdV hierarchy. In fact, we can 
now recognize {\it a posteriori} the differential equation of \fs\ as 
the reduction of the KdV equation, whose quasi-periodic solutions are of
course elliptic  functions. It is interesting to notice that the differential
equations governing the blowup behavior of $SU(N)$ topological Yang-Mills
theory in four dimensions  turn out to be essentially the same than the
equations governing the correlation functions of two-dimensional topological
gravity \wittopgrav. This is yet another manifestation of the  intimate
relationship between 4d ${\cal N}=2$ theories and 2d physics.    

\newsec{Manifolds of simple type and multisoliton solutions}

\subsec{${\cal N}=1$ points}

There are points in the moduli space of the hyperelliptic curve where one has
maximal degeneration, {\it i.e.}, all the $B_i$ cycles collapse. These
points are usually called, in the context of ${\cal N}=2$ gauge  theories,
the ${\cal N}=1$ points, since these are the confining  vacua that one 
obtains after breaking ${\cal N}=2$ down to ${\cal N}=1$. The physics of 
these points in pure Yang-Mills theory has been studied in detail in 
\ds, and some aspects have been addressed in \edemas\ from the point 
of view of the Whitham hierarchy. In this subsection we will rederive some of
the  results of \ds\edemas\ by using the approach of \belo, section 4.4. 
In particular, we will obtain a compact expression for the leading 
contribution of the off-diagonal magnetic couplings near the ${\cal N}=1$ 
points. 

The ${\cal N}=1$ points of the ${\cal N}=2$ gauge theory are described 
by Chebyshev polynomials. The polynomial $P_N(x)$ becomes \foot{We set for
convenience $\Lambda = 1$ along this section.}, at a point of  maximal
degeneration, 
\eqn\cheby{
P_N(x) =2 \cos \bigl( N \arccos {x \over 2} \bigr) ~,}
and the other ${\cal N}=1$ points are obtained using the ${\IZ}_N$
symmetry of the theory. From now on we will focus on the ${\cal N}=1$ point
corresponding to \cheby. The branch points of the curve are now the single 
branch points $e_1=-e_{2g+2}=2$, and the double branch points are: 
\eqn\brunch{
e_{2k}=e_{2k+1}=\widehat \phi_k = 2 \cos {\pi k \over N} ~, \,\,\,\,\,\,\,\,\
k=1, \cdots, g ~.} 
The values of the Casimirs at this degeneration are given by the elementary
symmetric polynomials of the eigenvalues $2 \cos {\pi (k-1/2) \over N}$,~
$k=1, \cdots, N$ \ds. For example, 
\eqn\simpleu{
u_2 = N, \,\,\,\,\ u_3=0, \,\,\,\,\ u_4={N\over 2}(3-N) ~.}  
When the curve degenerates in the way specified by \cheby, the $B_i$ cycles
surround the points $\widehat \phi_i$ clockwise, while the $A^i$ cycles
become curves going from $\widehat \phi_i$ to $2$ on the upper sheet and
returning to $\widehat \phi_i$ on the lower sheet. The hyperelliptic curve
\hyper\ becomes 
\eqn\limitcur{
y ={\sqrt {x^2-4}} ~\prod_{k=1}^g (x-\widehat \phi_k) ~.}
Consider now the normalized ``magnetic'' holomorphic differentials: 
\eqn\normhol{
\omega^i = (B^{-1})^{k i}dv_k ={\varphi^i (x) dx \over y} ~.}
Then, it follows from \periods\ that 
\eqn\residuo{
{1\over 2\pi i} \oint_{B_j} \omega^i = -{\rm res}_{x =\widehat \phi_j} 
\omega^i=\delta^i_j ~.}
Using the explicit expressions \holo\ and \limitcur, we find:
\eqn\laphi{
\varphi^j(x)=-2i \sin{ \pi j\over N} \prod_{l\not= j} (x-\widehat \phi_l) ~,} 
and 
\eqn\norholdef{
\omega^j=-{2i \sin{ \pi j\over N} \over {\sqrt {x^2-4}} ~(x-\widehat
\phi_j) ~.}}
Let $S_0=1$, $S_j=\sum_{i_1 < \cdots < i_j} x_{i_1} \cdots x_{i_j}$
be the  elementary symmetric polynomial of degree $j$. From \normhol, \laphi\
and \identif\  one deduces:
\eqn\ders{
{\partial u_{\ell+1} \over 
\partial a_{D,m}}= 2i(-1)^{\ell} \sin{ \pi m\over N} ~
S_{\ell-1} (\widehat \phi_{p\not=m}) ~.}
One can in fact check that this expression agrees with the results of \ds. 
Indeed, one can rederive from \ders\ equation (5.3) of \emm. 

Near the ${\cal N}=1$ points, the diagonal components of the ``magnetic''
couplings diverge, but the off-diagonal components are finite. The leading
terms of the off-diagonal components have been  investigated in \ds, 
where an implicit expression for them was proposed in terms of an integral 
involving a scaling trajectory. In \edemas\ it was shown that the 
Whitham hierarchy gives some nontrivial constraints on these terms, and an 
explicit expression satisfying the constraints was proposed. We will now 
derive a very simple expression for the leading terms of the off-diagonal 
couplings. From the above considerations it follows that 
\eqn\magco{
\tau_D^{k \ell}={1\over \pi i} \int_{\widehat \phi_k}^2 \omega^\ell ~.}
Taking into account \norholdef, the computation of \magco\ reduces to 
an elementary integral \belo. Denoting: 
\eqn\gas{
\gamma_j =-i {\sqrt {\widehat \phi_j -2 \over \widehat \phi_j +2}}= 
\tan {\pi j \over 2 N} ~,}
we find
\eqn\offdiag{
\tau_D^{k\ell} ={1 \over \pi i} \log {\gamma_\ell -\gamma_k \over
\gamma_\ell+\gamma_k} ~, \,\,\,\,\,\,\,\,\ k<\ell ~.} 
We have checked that this expression agrees with the proposal of \edemas\ up
to $N=5$, although \offdiag\ is considerably simpler. Finally, notice that
the diagonal couplings diverge logarithmically  $\tau_D^{ii}\rightarrow
i\infty$ \ds. 
 
\subsec{The blowup function for manifolds of simple type}

We are now ready to compute the blowup function for manifolds of simple type.
The first thing we have to do is to rewrite \blowupf\ in  the magnetic frame
which is appropriate to the strong coupling regime, as in the related
analysis of \edemas.  Since the blowup function is invariant under duality
transformations,  the only change will be in the characteristic, which is now
$[\vec 0, \vec\Delta]$, and in the substitution of all the variables by
their duals  ({\it i.e.} we will have $\tau_D^{ij}$ instead of $\tau_{ij}$,
and $\partial u_{k}/\partial a_{D,i}$ instead of $\partial u_{k}/\partial
a^i$.) 

We now have all the ingredients to investigate the blowup function 
for manifolds of simple type. The dual theta function $\Theta_D[\vec 0, 
\vec\Delta](\vec \xi |\tau)$ vanishes at the ${\cal N}=1$ point, but after 
quotienting by $\Theta_D[\vec 0, \vec\Delta](0|\tau)$ we get a 
finite result:
\eqn\blowup{
{1\over C}\sum_{s_j=\pm 1} \prod_{p<q} \Bigl( {\gamma_q -\gamma_p \over
\gamma_q + \gamma_p}\Bigr)^{s_p s_q/2} \exp \Bigl\{ \sum_{l=2}^N{{i s_j t_l
\over 2} {\partial u_{l} \over \partial  a_{D,j}}}\Bigr\} ~,} 
where 
\eqn\acero{
C=\sum_{s_p=\pm 1} \prod_{p<q} \Bigl( {\gamma_q -\gamma_p 
\over \gamma_q + \gamma_p}\Bigr)^{s_p s_q/2} ~,}
and the values of the $B$-periods at the ${\cal N}=1$ points are given in 
\ders. To derive the above equation, we have used the explicit expression for
the offdiagonal couplings \offdiag. The values of the contact terms can be
obtained from the logarithmic derivatives of \blowup\  following \confin, but
it proves to be much more useful to use our  new equation for the contact
terms \cont. We just have to  compute the $B$-periods of the Abelian
differentials \drzero\ at the ${\cal N}=1$ point. This is easy to do by
making the change of variables $x=2 \cos \theta$ \ds. One has 
\eqn\abnone{
dr^\ell=i P'_\ell(\theta) \cot N\theta ~\sin \theta ~d\theta~,} 
with periods:
\eqn\peres{
\eta^\ell_{\,\, k}={\rm res}_{\theta =\hat \theta_k} dr^{\ell} = {i \over
N}P'_\ell(\widehat \phi_k) \sin {k  \pi \over N} ~,} 
where $\hat \theta_k = k\pi/N$. The contact terms are then given by: 
\eqn\contnone{
\CT_{k,\ell}= {i \over 2N} P'_{k-1}(\widehat \phi_m) \sin {m \pi \over N}
~{\partial u_{\ell} \over \partial  a_{D,m}} ~.} 
One has, for example:
\eqn\contnonebis{
\eqalign{
\CT_{2,\ell}=&{i \over 2N}\sin {m \pi \over N}{\partial u_{\ell} \over
\partial  a_{D,m}}={\ell \over 4N}u_{\ell} ~, \cr
\CT_{3,\ell}=&
{i \over N}\sin {2 m \pi \over N}{\partial u_{\ell} \over \partial 
a_{D,m}} ~.\cr}}
We have checked for low values of $N$ that the expression \contnone\ agrees
with the one obtained using \confin. Clearly, \contnone\ is much more compact
in this case. The last expression in the first line  of \contnonebis\
actually follows from \contu, but can be checked using the  results of this
section.  

Putting all the ingredients together, we find that the blowup function at
the ${\cal N}=1$ point is given by
\eqn\finalblow{
\eqalign{
\tau (t_i) = & ~{1\over C}\exp \Biggl\{ -\sum_{k,\ell} t_{k}t_\ell {i \over
2N} P'_{k-1}(\widehat \phi_m) \sin {m \pi \over N}{\partial u_{\ell} \over
\partial  a_{D,m}} \Biggr\} \cr 
& ~~~~~~~\cdot \sum_{s_j=\pm 1} \prod_{p<q} \Bigl( {\gamma_q -\gamma_p \over
\gamma_q + \gamma_p}\Bigr)^{s_p s_q/2} \exp \Bigl\{ \sum_{l=2}^N{{i s_j t_l
\over 2} {\partial u_{l} \over \partial  a_{D,j}}}\Bigr\} ~.\cr}} 
From the point of view of the underlying KdV hierarchy, this blowup function
has a very simple interpretation: it is a $\tau$ function for an
$(N-1)$-soliton solution, after making the linear transformation of times
explained in section 5. This is a simple consequence of the  fact that
quasi-periodic solutions of the KdV hierarchy become  multisoliton solutions
in the limit of maximal degeneracy of the  underlying Riemann surface (see,
for example, \belo\mum).   

An important consistency check of \finalblow\ can be made by considering the
explicit expression of the  Donaldson-Witten generating function for 
manifolds $X$ of simple type with $b_2^+(X) > 1$ obtained in
\mmtwo, which  is trivially extended to include more general descent
operators: 
\eqn\sundw{
\eqalign{
Z(p_k, f_k,S)_X^{{\cal N}=1}=& ~\alpha^{\chi}\beta^{\sigma} \sum_{x_j}\bigl(
\prod_{j=1}^{N-1} SW(x_j) \bigr) \prod_{j<k} \Bigl( {\gamma_k -\gamma_j \over 
\gamma_j + \gamma_k} \Bigr)^{-{(x_j,x_k)/ 2}} \cr
& ~~~~~~~\cdot \exp\Bigl\{ \sum_{k=2}^N \Bigl( p_k u_k -{i\over 2}f_k
{\partial u_{k} \over \partial a_{D,j}}(S,x_j)\Bigr) + S^2 \sum_{k,l}f_kf_l
\CT_{k,l}  \Bigr\} ~.\cr}}
In this equation, we have only recorded the contribution of one of the 
${\cal N}=1$ points, since the contributions of the other points follow 
from ${\IZ}_N$ symmetry. For each $i=1, \cdots, N-1$, the sum over $x_i$ 
is over all the Seiberg-Witten basic classes of the manifold $X$ \monopole,
whose Seiberg-Witten invariants are denoted by $SW(x_j)$. The values of the
$B$-periods and the contact terms are  those given in \ders\ and \contnone,
respectively. $(\,\  , \,\ )$  denotes the product in (co)homology. Finally,
$\alpha$ and $\beta$ are universal constants that only depend on $N$.  
If we now perform a blowup, for every basic class $x$ of $X$ we will 
obtain the basic classes $x \pm B$ in $\widehat X$, where $x$ denotes 
the pullback to $\widehat X$ of the basic class of $X$ \fssw. The
Seiberg-Witten  invarians are $SW(x\pm B) =SW(x)$ \fssw. If we now consider 
$Z(p_k, f_k,S)_{\widehat X}^{{\cal N}=1}$, we will have to substitute 
$x_i \rightarrow x_i +s_iB$ in \sundw, with $s_i =\pm 1$. The sum over basic
classes of the blownup  manifold $\widehat X$ factorizes into a sum over the
$x_i$ and  a sum over the $s_i$. Taking into account that $(x,B)=0$ for 
any cohomology class $x$ pulled back from $X$ to the blownup manifold, 
and that $B^2=-1$, we find that, under blowup, \sundw\ gets an extra factor
which exactly  agrees with \finalblow\ up to an overall constant \foot{The
overall  normalization also agrees if one takes into account the universal 
constants of the $u$-plane integral in the definition of the blowup
function.}. This is an important consistency check of the whole story and in
particular of the expression \sundw. The check is not trivial since, when
using \sundw, we have to rely on properties of the Seiberg-Witten invariants,
while \finalblow\ was derived by means of the $u$-plane integral.  

Let us finish this section by considering in detail the expression we
obtained for the blowup function at the ${\cal N}=1$ point \finalblow\ for
the cases of lower genus. For $g=2$, for example, $\gamma_1=1/{\sqrt 3}$ and
$\gamma_2={\sqrt 3}$. After using the explicit values of the $B$-periods
given in \ders, we obtained:
\eqn\suthreeblow{
\tau_{SU(3)}(t_2,t_3)={1\over 3}{\rm e}^{-{1\over 2}t_2^2 -t_3^2} 
\Bigl\{ \cosh ( {\sqrt 3}t_2) + 2 \cosh ( {\sqrt 3}t_3) \Bigr\} ~.}
Notice that the blowup function for simple type manifolds is given by a
compact expression as \suthreeblow, in contrast to the case of non-simple
type manifolds that we analyzed above. This fact was already observed in the
elliptic case \fs, and is related to the degeneration of hyperelliptic
functions to trigonometric functions. On the other hand, both expressions
must coincide as long as the blowup function is duality invariant. This means
that the whole expansion \expansion\ must reorganize itself into \suthreeblow\
when $u_2 = 3$ and $u_3 = 0$. Indeed, in expanding \suthreeblow\ up to sixth
order in the times
\eqn\expsimple{
\tau_{SU(3)}(t_2,t_3) = 1-{1\over 2}t_2^2t_3^2 -{1\over 4}t_3^4 -{1\over
120}t_2^6 +{1\over 8}t_2^4t_3^2 +{3\over 8}t_2^2t_3^4 +{13\over 120}t_3^6 +
\cdots ~,}
we find complete agreement with the expansion \sigdos\ in the nonsimple type
case. This is an important consistency check of the results of this paper.

For $g=3$, one has $\gamma_1={\sqrt 2}-1$, $\gamma_2=1$ and $\gamma_3={\sqrt
2}+1$, and the blowup function turns out to be:
\eqn\sufourblow{
\eqalign{
\tau_{SU(4)}(t_2,t_3,t_4) = & ~{1\over 4{\sqrt 2}}{\rm e}^{-{1\over 2}t_2^2
-t_3^2-2t_4^2 +t_2t_4} \Bigl\{ {\sqrt 2} \cosh ( t_2+2t_3-2t_4) \cr 
& + {\sqrt 2} \cosh ( -t_2+2t_3+2t_4) + ({\sqrt 2}-1) \cosh (({\sqrt
2}+1)t_2+2t_4) \cr
& + ({\sqrt 2}+1) \cosh ( ({\sqrt 2}-1)t_2-2t_4) \Bigr\} ~.}} 
Again, it is immediate to check that the leading terms of its expansion, 
\eqn\expsimple{
\tau_{SU(4)}(t_2,t_3,t_4) = 1-{1\over 6}t_3^4 -t_2t_3^2t_4 - {1\over
2}t_2^2t_4^2  +{4\over 3}t_2t_4^3 -{4\over 3}t_4^4 +\cdots ~,}
are in agreement with the result obtained in the nonsimple type case, after
taking into account that $u_2=4$ and $u_3=0$ at the ${\cal N}=1$ point. 

\newsec{Concluding Remarks}

In this paper we have carried out a detailed analysis of blowup formulae in
$SU(N)$ Donaldson-Witten theory. In particular, we have found an explicit
procedure to expand it in terms  of the Casimirs of the gauge group up to
arbitrary order, by using the theory of hyperelliptic Kleinian functions.
This theory clarifies in fact many other aspects of  blowup formulae and the
$u$-plane integral, like contact terms and the  relation with integrable
hierarchies. 

Although higher rank generalizations of Donaldson-Witten theory seem to be
rather intractable mathematically, it is likely that the behavior of the
higher rank invariants under blowup can be  determined by using only a
limited amount of information, like in the work of Fintushel and Stern \fs.
This article gives very precise  predictions for this behavior. In particular,
it implies that  the higher rank generalization of the differential equations 
studied in \fs\ will be essentially the KdV hierarchy. 

Our work can be generalized in many different directions. First of all, we
have analyzed only the case of $\vec\beta=\vec 0$, and certainly this is only
one particular case of the general blowup formula.  More work is needed along
this direction. In particular, it would require a generalization of the
procedure developed in section 3 for other kind of $\sigma$-functions

It would be also interesting to work out the details for theories including
massive hypermultiplets and/or other gauge groups. One of the most
interesting aspects of the theories with matter is that the magnetic flux
turns out to be fixed by topological constraints, and this gives a nonzero
value of $\vec\beta$ in the blowup function \mw\egm. 

Another direction to explore is the relation between the hyperelliptic
Kleinian functions and the theory of the prepotential. The blowup function
gives a natural set of Abelian differentials of the second kind, and we know
from general principles that such a set is one of the basic ingredients in
the construction of a Whitham hierarchy \Krichever. It would be very
interesting to develop this relation in general, at least for hierarchies
associated to hyperelliptic curves. This would further clarify the 
relations between blowup functions in generalizations of Donaldson-Witten
theory, and the construction  of Whitham hierarchies for supersymmetric
${\cal N}=2$ theories in \naka\itomoro\ITEP\emm\egrmm.

\bigskip
\noindent
{\bf Acknowledgements:} We would like to thank Javier Mas 
for useful discussions and for a critical reading of the 
manuscript. We are specially indebted to Victor Enolskii for 
sharing with us his knowledge of the theory of Kleinian 
functions and for his patient 
explanations. The work of J.D.E. has been supported by the National Research
Council (CONICET) of Argentina. The work of M.M. has been 
supported by DOE grant DE-FG02-96ER40959.  

\listrefs

\bye